\documentclass[12pt]{article}   	
\usepackage{geometry}                		
\usepackage{graphicx}				
\usepackage{amssymb}
\usepackage{amsmath}
\usepackage{epstopdf}
\usepackage{natbib}
\usepackage{hyperref}
\usepackage{setspace}
\usepackage{afterpage}
\usepackage{authblk}
\hypersetup{colorlinks=true, linkcolor=blue, citecolor=blue,urlcolor=blue}
\usepackage[all]{hypcap}
\usepackage[normalem]{ulem}  

\def\eqlbl#1{\label{eq:#1}}
\def\eqref#1{(\ref{eq:#1})}

\title{Whole planet coupling between climate, mantle, and core: Implications for the evolution of rocky planets }
\author{Bradford J. Foley, Peter E. Driscoll}
\affil{Department of Terrestrial Magnetism, Carnegie Institution for Science, Washington, DC}
\date{}	

\begin{document}
\maketitle

\begin{abstract}
Earth's climate, mantle, and core interact over geologic timescales. Climate influences whether plate tectonics can take place on a planet, with cool climates being favorable for plate tectonics because they enhance stresses in the lithosphere, suppress plate boundary annealing, and promote hydration and weakening of the lithosphere. Plate tectonics plays a vital role in the long-term carbon cycle, which helps to maintain a temperate climate. Plate tectonics provides long-term cooling of the core, which is vital for generating a magnetic field, and the magnetic field is capable of shielding atmospheric volatiles from the solar wind. Coupling between climate, mantle, and core can potentially explain the divergent evolution of Earth and Venus. As Venus lies too close to the sun for liquid water to exist, there is no long-term carbon cycle and thus an extremely hot climate. Therefore plate tectonics cannot operate and a long-lived core dynamo cannot be sustained due to insufficient core cooling. On planets within the habitable zone where liquid water is possible, a wide range of evolutionary scenarios can take place depending on initial atmospheric composition, bulk volatile content, or the timing of when plate tectonics initiates, among other factors. Many of these evolutionary trajectories would render the planet uninhabitable. However, there is still significant uncertainty over the nature of the coupling between climate, mantle, and core. Future work is needed to constrain potential evolutionary scenarios and the likelihood of an Earth-like evolution.   
\end{abstract}

\section{Introduction}
\subsection{Overview}  

Recent discoveries have revealed that rocky exoplanets are relatively common \citep{Batalha2014}. As a consequence, determining the factors necessary for a rocky planet to support life, especially life that may be remotely observable, has become an increasingly important topic. A major requirement, that has been extensively studied, is that solar luminosity must be neither too high nor too low for liquid water to be stable on a planet's surface; this requirement leads to the concept of the ``habitable zone," the range of orbital distances where liquid water is possible \citep{Hart1978,Hart1979,Kasting1993,Franck2000,Kopp2014}. Inward of the habitable zone's inner edge, the critical solar flux that triggers a runaway greenhouse effect is exceeded. The critical flux is typically estimated at $\approx 300 $ W m$^{-2}$, with variations of $\sim 10-100$ W m$^{-2}$ possible due to atmospheric composition, planet size, or surface water inventory \citep{Ingersoll1969,Kasting1988,Nakajima1992,Abe2011,Goldblatt2013}. In a runaway greenhouse state liquid water can not condense out of the atmosphere, so any water present would exist as steam. Furthermore a runaway greenhouse is thought to cause rapid water loss to space, and can thus leave a planet desiccated \citep{Kasting1988,Hamano2013,Wordsworth2013}. Beyond the outer edge insolation levels are so low that no amount of CO$_2$ can keep surface temperatures above freezing \citep{Kasting1993}. 

However, lying within the habitable zone does not guarantee that surface conditions will be suitable for life. Variations in atmospheric CO$_2$ content can lead to cold climates where a planet is globally glaciated, or hot climates where surface temperatures are higher than any known life can tolerate (i.e. above $\approx 400$ K \citep{Takai2008}). A hot CO$_2$ greenhouse can also cause rapid water loss to space \citep{Kasting1988} (though \cite{Wordsworth2013} argues against this), or even a steam atmosphere if surface temperatures exceed water's critical temperature of 647 K. Moreover, the solar wind can strip the atmosphere of water and expose the surface to harmful radiation, unless a magnetic field is present to shield the planet \citep[e.g.][]{Kasting2003,griessmeier2009,brain2014}. 

Atmospheric CO$_2$ concentrations are regulated by the long-term carbon cycle on Earth, such that surface temperatures have remained temperate throughout geologic time \citep[e.g.][]{walker1981,Berner2004}. The long-term carbon cycle is facilitated by plate tectonics \citep[e.g.][]{Kasting2003}. Furthermore the magnetic field is maintained by convection in Earth's liquid iron outer core (i.e. the geodynamo). As a result, interior processes, namely the operation of plate tectonics and the geodynamo, are vital for habitability. However, whether plate tectonics or a strong magnetic field is likely on rocky planets, especially those in the habitable zone where liquid water is possible, is unclear. The four rocky planets of our solar system have taken dramatically different evolutionary paths, with only Earth developing into a habitable planet possessing liquid water oceans, plate tectonics, and a strong, internally generated magnetic field. In particular the contrast between Earth and Venus, which is approximately the same size as Earth and has a similar composition yet lacks plate tectonics, a magnetic field, and a temperate climate, is striking. 

In this review we synthesize recent work to highlight that plate tectonics, climate, and the geodynamo are coupled, and that this ``whole planet coupling" between surface and interior places new constraints on whether plate tectonics, temperate climates, and magnetic fields can develop on a rocky planet. We hypothesize that whole planet coupling can potentially explain the Earth-Venus dichotomy, as it allows two otherwise similar planets to undergo drastically different evolutions, due solely to one lying inward of the habitable zone's inner edge and the other within the habitable zone. We also hypothesize that whole planet coupling can lead to a number of different evolutionary scenarios for habitable zone planets, many of which would be detrimental for life, based on initial atmospheric composition, planetary volatile content, and other factors. We primarily focus on habitable zone planets, as these are most interesting in terms of astrobiology, and because the full series of surface-interior interactions we describe involves the long-term carbon cycle, which requires liquid water. Each process, the generation of plate tectonics from mantle convection, climate regulation due to the long-term carbon cycle, and dynamo action in the core, is still incompletely understood and the couplings between these processes are even more uncertain. As a result, significant future work will be needed to place more quantitative constraints on the evolutionary scenarios discussed in this review. 

\subsection{Whole planet coupling}  

\begin{figure}[ht]
\includegraphics[width=\linewidth]{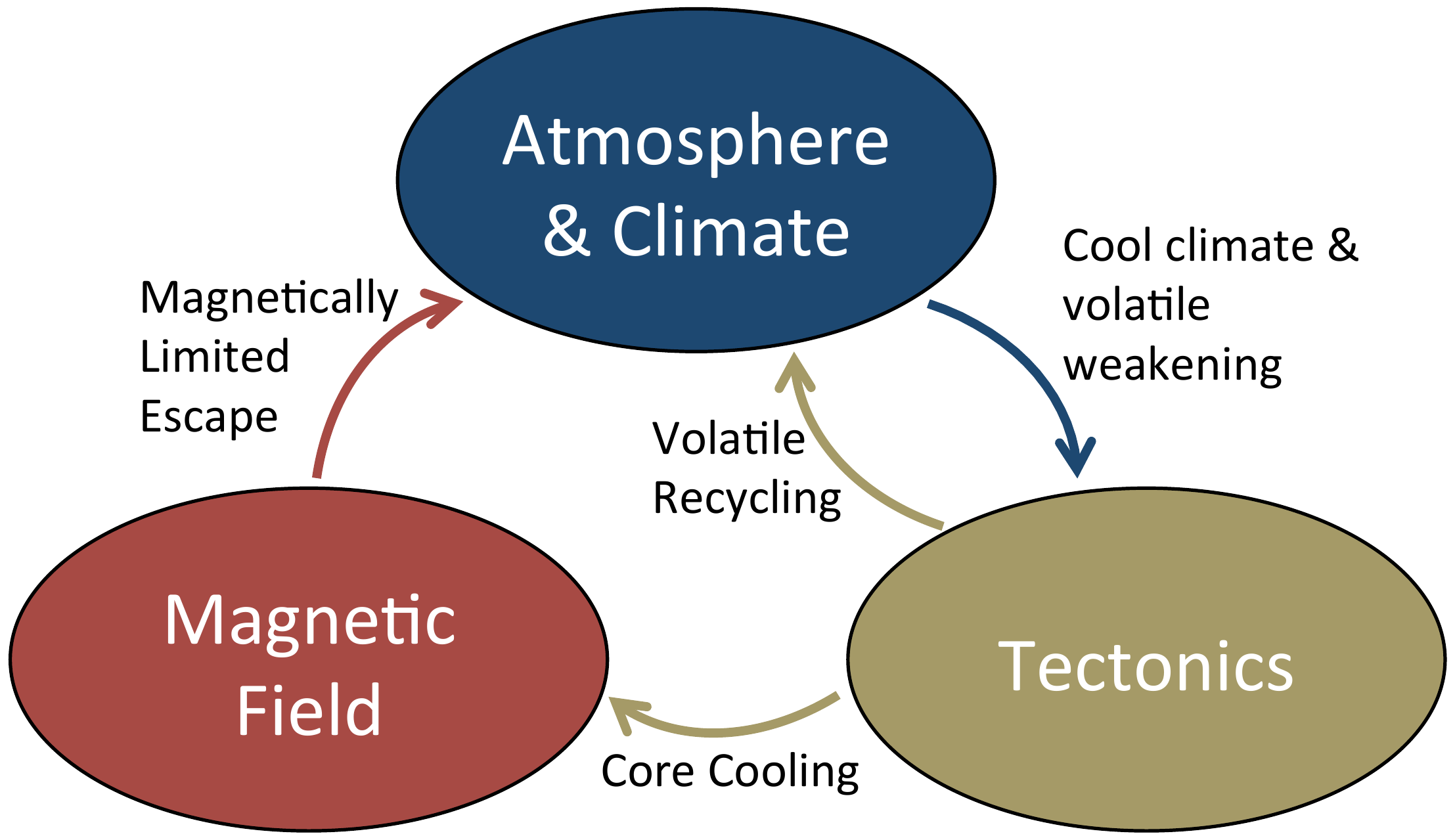}
\caption{Flow chart representing the concept of whole planet coupling. Climate influences tectonics through the role of surface temperature in a planet's tectonic regime (i.e. stagnant lid versus plate tectonics), while the tectonic regime in turn affects climate through volatile cycling between the surface and interior. The tectonic regime also influences whether a magnetic field can be generated by dictating the core cooling rate.  Finally, the strength of the magnetic field influences atmospheric escape, and therefore long-term climate evolution. }
\label{fig:CTM}
\end{figure}

Several basic concepts illustrate the coupling between the surface and interior (Figure \ref{fig:CTM}). (1) Climate influences whether plate tectonics can take place on a planet. (2) Plate tectonics plays a vital role in the long-term carbon cycle, which helps to maintain a temperate climate. (3) Plate tectonics effects the generation of the magnetic field via core cooling. (4) The magnetic field is capable of shielding the atmosphere from the solar wind. Cool climates are favorable for plate tectonics because they facilitate the formation of weak lithospheric shear zones, which are necessary for plate tectonics to operate. Low surface temperatures suppress rock annealing, increase the negative buoyancy of the lithosphere, and allow for deep cracking and subsequent water ingestion into the lithosphere, all of which promote the formation of weak shear zones. When liquid water is present on a planet's surface, silicate weathering, the primary sink of atmospheric CO$_2$ and thus a key component of the long-term carbon cycle, is active. However, silicate weathering also requires a sufficient supply of fresh, weatherable rock at the surface, which plate tectonics helps to provide via orogeny and uplift. As a result, the coupling between plate tectonics and climate can behave as a negative feedback mechanism in some cases, where a cool climate promotes the operation of plate tectonics, and plate tectonics enhances silicate weathering such that the carbon cycle can sustain cool climate conditions. 

An additional coupling comes into play via the core dynamo and the magnetic field. The magnetic field is generated by either thermal or chemical convection in the liquid iron core. Thermal convection requires a super-adiabatic heat flux out of the core, which is controlled in part by the style of mantle convection, while chemical convection is driven by light element release during inner core nucleation, which also relies on cooling of the core. Plate tectonics cools the mantle efficiently by continuously subducting cold slabs into the deep interior, thus maintaining a high heat flow out of the core. The magnetic field can in turn limit atmospheric escape, helping retain liquid surface water.  The coupling between plate tectonics and the core dynamo, and the magnetic field and the climate, completes the concept of whole planet coupling (Figure \ref{fig:CTM}). Moreover the magnetic field and plate tectonics can also act as a negative feedback in cases where magnetic shielding is required to prevent rapid planetary water loss, and plate tectonics is needed to drive the dynamo.  

\subsection{Applications to the evolution of rocky exoplanets}  

The concept of whole planet coupling discussed in this review is based on our knowledge of the Earth and the other rocky planets in the solar system, and is therefore limited to planets of a particular composition; i.e. planets made up mainly of silicate mantles and iron cores. It is unknown whether any of the processes discussed here are applicable to planets with more exotic compositions, such as planets primarily composed of carbides rather than silicates \citep[e.g.][]{Madhu2012}. Furthermore we focus on H$_2$O and CO$_2$ as key volatiles; CO$_2$ is an important greenhouse gas for stabilizing planetary climate, and water is crucial for driving the carbon cycle, may play a role in the operation of plate tectonics, and is thought to be a necessary ingredient for life. Thus planets must accrete a significant supply of both H$_2$O and CO$_2$ for whole planet coupling to be possible. However, volatile accretion is unlikely to be a problem as simulations indicate that planets commonly acquire large volatile inventories (i.e. Earth-like or larger) during late stage planet formation \citep{Morbi2000,Raymond2004}. 

Another important consideration is the redox state of the mantle, which determines whether degassing via planetary magmatism releases H$_2$O and CO$_2$ to the atmosphere or reduced species such as H$_2$ and CH$_4$ \citep[e.g.][]{Kasting1993b}. Carbon dioxide is the only major greenhouse gas known to be regulated by negative feedbacks such that it has a stabilizing influence on climate.  Thus having CO$_2$ as a primary greenhouse gas is important for the whole planet coupling discussed here to operate. An oxidized upper mantle favors CO$_2$ over CH$_4$ and other reduced gases. Earth's mantle has been oxidized at present day levels since at least the early Archean \citep{Delano2001}, and possibly even since the Hadean \citep{Trail2011}. Oxidation of the mantle is thought to occur by disproportionation of FeO to Fe$_2$O$_3$-bearing perovskite and iron metal in the lower mantle during accretion and core formation. The iron metal is then lost to the core leaving behind oxidized perovskite that mixes with the rest of the mantle \citep{Frost2008,Frost2008_rev}. Disproportionation of FeO is expected to occur on rocky planets Earth sized or larger \citep{Wade2005,Wood2006}, so CO$_2$ is likely to be an important greenhouse gas on exoplanets. Though other greenhouse gases can still be important, any planet where significant amounts of CO$_2$ are degassed by mantle volcanism will need silicate weathering to act as a CO$_2$ sink to avoid extremely hot climates.  

\subsection{Outline}
The paper is structured as follows. We review the basic physics behind the generation of plate tectonics from mantle convection, highlighting the role of climate, in \S \ref{sec:plate_generation}. We then review the long-term carbon cycle and its influence on climate, and detail the specific ways plate tectonics is important for the operation of this cycle, in \S \ref{sec:climate_reg_cc}. Next we review magnetic field generation via a core dynamo and the physics of atmospheric shielding and volatile retention in \S \ref{sec:whole_planet}. In \S \ref{sec:summary} we integrate the discussion of the previous  three sections to describe the range of different evolutionary scenarios that can result from differences in orbital distance, initial climate state, volatile inventory, and other factors as a consequence of whole planet coupling. We conclude by listing some major open questions that must be addressed in order to further our understanding of how terrestrial planets evolve and the geophysical factors that influence habitability (\S \ref{sec:future}).  

\section{Generation of plate tectonics from mantle convection and the role of climate}
\label{sec:plate_generation}
     
\subsection{General physics of plate generation}

\label{sec:plate_physics_gen}

Plate tectonics is the surface expression of convection in the Earth's mantle: the lithosphere, composed of the individual plates, is the cold upper thermal boundary layer of the convecting mantle, subducting slabs are convective downwellings, and plumes are convective upwellings \citep{davies1999,Berco2015_treatise}.  However, mantle convection does not always lead to plate tectonics, as evidenced by Mercury, Mars, and Venus, each of which is thought to have a convective mantle but lack plate tectonics \citep[e.g.][]{Breuer2007}. On these planets, ``stagnant lid convection," where the lithosphere acts as a rigid, immobile lid lying above the convecting mantle \citep[e.g.][]{ogawa1991,Davaille1993,slava1995}, is thought to operate \citep{Strom1975,Phillips1981,Solomon1992,Solomatov1996,Solomatov1997,Spohn2001,ONeill2007c}, though Venus may experience occasional, short-lived episodes of subduction \citep{Turcotte1993}. Stagnant lid convection is a result of the temperature dependence of mantle viscosity, which increases by many orders of magnitude as temperature decreases from the hot conditions that prevail in the planetary interior to the much cooler temperatures that prevail at the surface \citep[e.g.][]{Karato1993,hirth2003}. When the viscosity at the surface is $\sim 10^3 - 10^4$ times that of the interior, the top thermal boundary layer is so viscous that it can no longer sink under its own weight and form subduction zones, and stagnant lid convection ensues. \citep[e.g.][]{Richter1983,christensen1984c,ogawa1991,Davaille1993,Moresi1995,slava1995}. Laboratory experiments indicate that the viscosity ratio between the surface and mantle interior for typical terrestrial planets, including Earth, is $\sim 10^{10}-10^{20}$, far larger than the threshold for stagnant lid convection. Stagnant lid convection is thus the ``natural" state for rocky planets, and additional processes are necessary for plate tectonics to operate on Earth. 

In order to generate plate-tectonic style mantle convection, rheological complexities capable of forming weak, localized shear zones in the high viscosity lithosphere are necessary. These weak shear zones (essentially plate boundaries) remove the strong viscous resistance to surface motion that temperature-dependent viscosity causes and thus allow subduction and plate motions to occur. We discuss plate boundary formation in terms of viscous weakening in the lithosphere because the majority of the lithosphere, including the region of peak strength in the mid-lithosphere, deforms via ductile or semibrittle/semiductile behavior; the lithosphere only ``breaks" in a brittle fashion at shallow depths of $< 10-20$ km \citep{Berco2015_treatise}. Many mechanisms have been proposed for producing the lithospheric weakening and shear localization necessary for plate-tectonic style mantle convection \citep[see reviews by][]{Tackley2000,Berco2003,bk2003b,Regenauer2003,Berco2015_treatise}; while we won't detail each of these mechanisms, they do all share some general principles. Namely, shear-thinning non-Newtonian rheologies, have been found to be successful at generating plate-like mantle convection.    

\begin{figure}
\includegraphics[scale = 0.63]{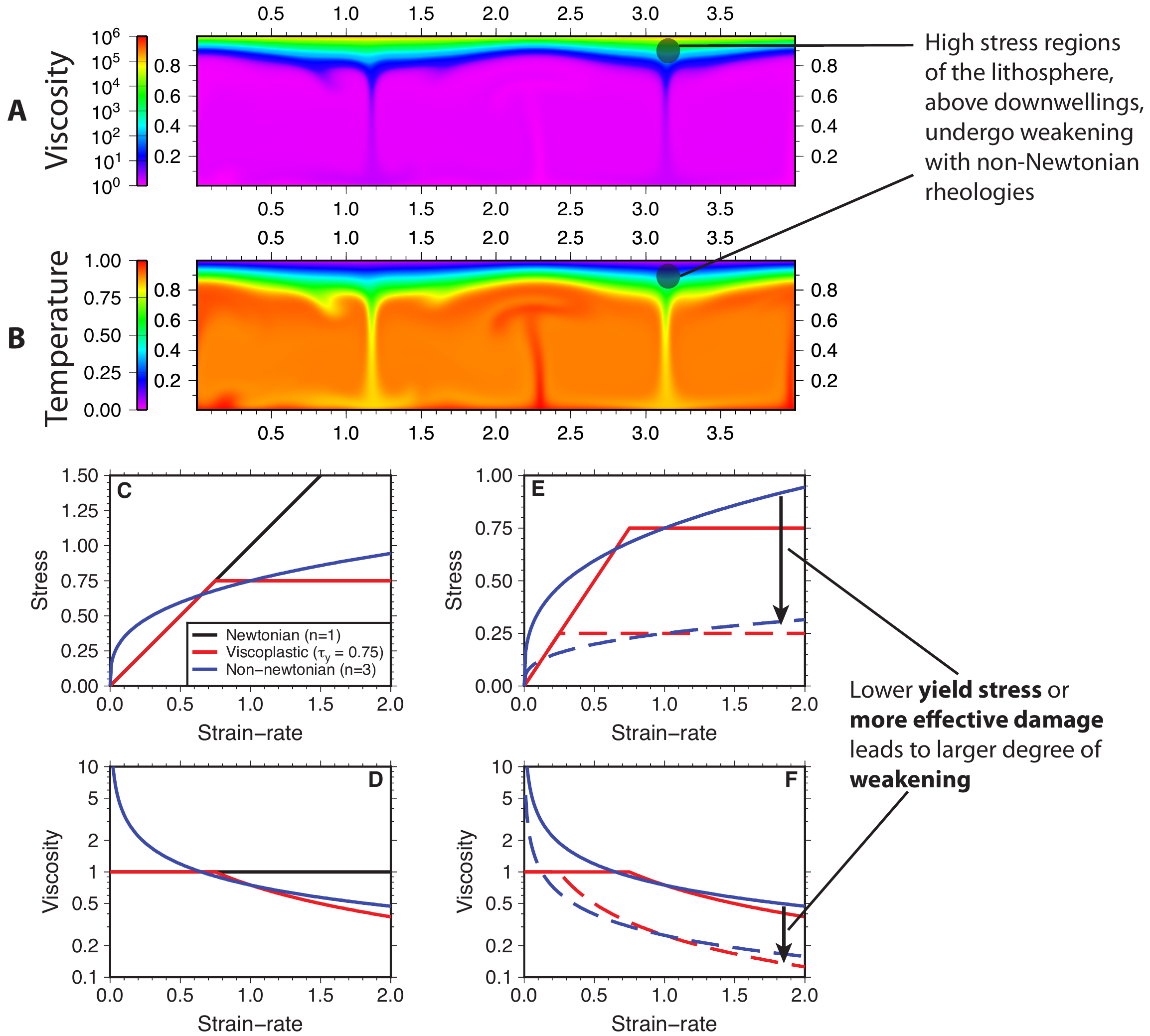}
\caption{\label {fig:plate_gen} Illustration of plate generation using non-Newtonian (or effectively non-Newtonian) rheologies. Shown are a typical viscosity field (A) and temperature field (B) for stagnant lid convection and schematic strain-rate versus stress (C,E) and viscosity versus strain-rate curves (D,F) for viscoplastic (see \S \ref{sec:plasticity}) and $n=3$ power law (equations \eqref{non_newt} \& \eqref{non_newt_mu}) rheologies. Weakening of the high viscosity lithosphere present during stagnant lid convection is needed in order to allow subduction, and with non-Newtonian rheologies such weakening will occur at high stress regions (highlighted in grey in panels (A) and (B)). Panels (C) and (D) illustrate how non-Newtonian rheologies cause viscous weakening at high stresses or strain-rates, while panels (E) and (F) schematically demonstrate how outside conditions that influence the effectiveness of a given plate generation mechanism change the strain-rate versus stress curve (E) and the viscosity versus strain-rate curve (F). }  
\end{figure}

In a shear-thinning fluid, the strain-rate, $\dot{\varepsilon}$, is a non-linear function of the stress, $\tau$, (where $\dot{\varepsilon}$ and $\tau$ are the second invariants of the stress and strain-rate tensors, respectively): 
\begin{equation}
\eqlbl{non_newt}
\dot{\varepsilon} \propto \tau^n 
\end{equation}
where $n$ is a constant. As the effective viscosity, $\mu_{eff}$, is proportional to $\tau \dot{\varepsilon}^{-1}$,  
\begin{equation}
\eqlbl{non_newt_mu}
\mu_{eff} \propto \tau^{(1-n)}  \propto \dot{\varepsilon}^{\frac{(1-n)}{n}} .
\end{equation}
With $n > 1$, viscosity decreases as stress or strain-rate increases and high stress, high strain-rate regions of the lithosphere, such as regions of compression or tension caused by drips breaking off the base of the lithosphere, are weakened (see Figure \ref{fig:plate_gen}). Thus lithospheric weakening occurs in exactly the regions necessary for facilitating subduction. Shear-thinning behavior can result from inherently non-Newtonian rheologies with $n > 1$, or more complicated rheologies that involve ``memory," where a time-evolving state variable, itself a function of stress and/or strain-rate, influences viscosity. In the case of memory, an effectively shear-thinning $\dot{\varepsilon}-\tau$ relationship can be calculated by taking the state variable in steady-state (as we show for grainsize reduction in \S \ref{sec:damage}). Moreover, rheologies where stress decreases at high strain-rates, resulting in even stronger weakening, are possible but beyond the scope of this paper. 

\subsubsection{Viscoplasticity}
\label{sec:plasticity}

One popular plate generation mechanism is the viscoplastic rheology, where the lithosphere ``fails" and becomes mobilized when it's intrinsic strength, the yield stress, is reached \citep{Fowler1993,Moresi1998}. Typical implementations of the viscoplastic rheology assume that the mantle behaves like a Newtonian fluid (following \eqref{non_newt} \& \eqref{non_newt_mu} with $n = 1$) until stress becomes equal to the yield stress, $\tau_y$ (see Figures \ref{fig:plate_gen}C \& D). At this point stress is independent of strain-rate, fixed at $\tau_y$, and the viscosity follows 
\begin{equation}
\mu_{eff} \propto \frac{\tau_y}{\dot{\varepsilon}}.
\end{equation}
When convective stresses in the lithosphere hit the yield stress, low viscosity shear zones at convergent and divergent margins can form, and plate-like mantle convection can be obtained in two-dimensional \citep{Moresi1998,richards2001,korenaga2010}, three-dimensional cartesian \citep{trompert1998,Tackley2000a,stein2004}, and three-dimensional spherical convection models \citep{vanHeck2008,Foley2009}. Convective stresses lower than the yield stress do not form weak plate boundaries, and thus result in stagnant lid convection.  

The convective stress, $\tau_m$, has typically been found to scale as
\begin{equation}
\eqlbl{tau_m}
\tau_m \propto \mu_m \dot{\varepsilon}_m
\end{equation}   
where $\mu_m$ and $\dot{\varepsilon}_m$ are the effective viscosity and strain-rate in the mantle, respectively \citep[e.g.][]{ONeill2007b}. The strain-rate can be estimated as $v_m/d$, where $v_m$ is convective velocity and $d$ is the thickness of the mantle. Velocity scales as \citep[e.g.][]{slava1995,turc1982} 
\begin{equation}
\eqlbl{v_m}
v_m \propto Ra^{\frac{2}{3}} ,
\end{equation}  
where the Rayleigh number ($Ra$) is defined as 
\begin{equation}
\eqlbl{Ra}
Ra = \frac{\rho g \alpha \Delta T d^3}{\kappa \mu_m}
\end{equation}
and describes the convective vigor of the system $(\rho$ is density, $g$ gravity, $\alpha$ thermal expansivity, $\Delta T$ the non-adiabatic temperature drop across the mantle, and $\kappa$ the thermal diffusivity). 

However, despite a large Rayleigh number on the order of $10^7 - 10^8$, typical convective stresses are only $\sim 1-100$ MPa \citep[e.g.][]{Tackley2000a,Solomatov2004,ONeill2007}, approximately an order of magnitude lower than the strength of the lithosphere based on laboratory experiments \citep[e.g.][]{kohlstedt1995}. Thus much work has focused on mechanisms for decreasing the lithospheric strength, usually involving water \citep[e.g.][]{Tozer1985,Lenardic1994,rege2001,vanderLee2008}, though some recent models have looked at mechanisms for increasing convective stresses \citep{Rolf2011,Hoink2012}. In \S \ref{sec:climate_tectonics} we describe how the possibility of water weakening links the operation of plate tectonics to climate. Moreover, climate also influences convective stresses, by altering $\Delta T$, providing another important link between climate and tectonics.  

\subsubsection{Damage and grainsize reduction}
\label{sec:damage}  

While viscoplasticity has shown some success at generating plate-like mantle convection, it fails to form localized strike-slip faults \citep{Tackley2000a,vanHeck2008,Foley2009}, a problem with all simple non-Newtonian rheologies \citep{berco1993}. Furthermore viscoplasticity is an instantaneous rheology: when stresses fall below the yield stress weakened lithosphere instantly regains its strength. On Earth, however, dormant weak zones are known to persist over geologic timescales, and can serve as nucleation points for new subduction zones \citep{Toth1998,gurnis2000}. An alternative mechanism, that allows for dormant weak zones (i.e. memory of past deformation) and is capable of generating strike-slip faults, is grainsize reduction \citep{bk2003b,br2012,br2013,br2014}. Grainsize reduction is commonly seen in field observations of exhumed lithospheric shear zones \citep{white1980,Drury1991,Jin1998,Warren2006,Skemer2010}, and experiments show that viscosity is proportional to grainsize when deformation is dominated by diffusion creep or grain boundary sliding \citep[e.g.][]{hirth2003}.  

One major problem for a grainsize reduction plate generation mechanism, though, is that grainsize reduction occurs by dynamic recrystallization, which takes place in the dislocation creep regime, while grainsize sensitive flow only occurs in the diffusion creep or grain boundary sliding regimes \citep[e.g.][]{Etheridge1979,DeBresser1998,Karato1993}. Thus, a feedback mechanism, where deformation causes grainsize reduction, which weakens the material and leads to more deformation, is apparently not possible \citep[e.g.][]{DeBresser2001}. Without such a feedback only modest weakening can occur. However, a new theory argues that when a secondary phase (e.g. pyroxene) is dispersed throughout the primary phase (e.g. olivine), a grainsize feedback is possible, because grainsize reduction can continue in the diffusion creep regime due the combined effects of damage to the interface between phases and Zener pinning (where the secondary phase blocks grain-growth of the primary phase) \citep{br2012}. 

Using a simplified version of the theory in \cite{br2012} \citep[see also][]{Foley2013_scaling}, the grainsize reduction mechanism can be formulated as: 
\begin{equation}
\eqlbl{eqd}
\frac{dA}{dt} = \frac{f_{A}}{\gamma}\Psi - h A^p
\end{equation}
where $A$ is the fineness (or inverse grainsize), $\gamma$ is surface tension, $h$ is the grain-growth (or healing) rate, $p$ is a constant, $\Psi$ is the deformational work, defined as  $\Psi = \varepsilon_{ij}\tau_{ij}$, where $\varepsilon_{ij}$ is the strain-rate tensor and $\tau_{ij}$ is the stress tensor, and $f_A$ is the fraction of deformational work that partitions into surface energy, thus driving grainsize reduction. The grain-growth rate is strongly temperature dependent, and is given by
\begin{equation}
\eqlbl{eqheal}
h = h_n \exp \left(\frac{-E_h}{R_g T} \right)     
\end{equation}
where $h_n$ is a constant, $E_h$ is the activation energy for grain-growth, $R_g$ is the universal gas constant, and $T$ is temperature.  The effective viscosity is
\begin{equation}
\eqlbl{eqv}
\mu_{eff} = \mu_n \exp \left(\frac{E_v}{R_g T} \right) \left(\frac{A}{A_0} \right)^{-m} = \mu(T)\left(\frac{A}{A_0} \right)^{-m}  ,
\end{equation}
where $\mu_n$ and $m$ are constants, $E_v$ is the activation energy for diffusion creep, and $A_0$ is the reference fineness. Taking \eqref{eqd} in steady-state, using the viscous constitutive law $\tau_{ij} = 2 \mu_{eff} \varepsilon_{ij}$, and the definition $\varepsilon_{ij} \varepsilon_{ij} = 2 \dot{\varepsilon}$, the fineness can be written as a function of strain-rate
\begin{equation}
\eqlbl{steady_state}
A = \left(\frac{4 f_{A} \mu(T)\dot{\varepsilon}^2 A_0^m}{\gamma h} \right)^{\frac{1}{p+m}}  .
\end{equation}
Combining \eqref{eqv} and \eqref{steady_state}, the viscosity is also a function of strain-rate, 
\begin{equation}
\mu_{eff} = \left(\mu(T)^p \left(\frac{4 f_{A} \dot{\varepsilon}^2}{\gamma h A_0^p}\right)^{-m} \right)^{\frac{1}{p+m}}
\end{equation} 
and using $\tau = 2 \mu_{eff} \dot{\varepsilon}$ the effective $\tau-\dot{\varepsilon}$ relationship is
\begin{equation}
\eqlbl{dam_const_law}
\dot{\varepsilon} = \frac{1}{2} \left(\mu(T)^{-p} \left(\frac{f_{A} }{\gamma h A_0^p} \right)^m \tau^{p+m} \right)^{\frac{1}{p-m}}. 
\end{equation}
With typical parameters of $m=2$ and $p=4$ \citep{hirth2003,br2012}, the grainsize reduction mechanism results in an effectively non-Newtonian rheology with $n=3$ (Figures \ref{fig:plate_gen}C \& D). Furthermore, as detailed in \S \ref{sec:climate_tectonics}, climate can have a significant impact on plate generation with this mechanism, because grain-growth is temperature-dependent. At high surface temperatures, and thus higher temperatures in the mid-lithosphere, faster grain-growth (i.e. higher $h$) impedes grainsize reduction. 

\subsection{Influence of climate on tectonic regime}
\label{sec:climate_tectonics}

Both the viscoplastic and grainsize reduction mechanisms for generating plate-tectonic style mantle convection are linked to climate. Although we treat these mechanisms separately, they are not mutually exclusive. The viscoplastic rheology is most relevant to brittle deformation in the upper lithosphere, while grainsize reduction provides viscous weakening in the lower lithosphere. As a result both mechanisms can operate simultaneously, and both may be important for generating plate tectonics.

Climate can influence wether plate-like convection occurs with a viscoplastic rheology in three main ways: first, climate dictates whether water can exist in contact with ocean lithosphere at the surface, such that high pore pressure \citep[e.g.][]{Hubbert1959,Sibson1977,Rice1992,Sleep1992} and weak, hydrous phases \citep{Escartin2001,Hilariet2007} can lubricate faults and lower the lithosphere's yield strength; second, climate influences whether deep cracking of the lithosphere can occur, which is potentially important for hydrating the mid- to lower-lithosphere; and third, climate influences convective stresses. The first point, whether water can interact with rocks at the surface, provides only a weak influence of climate on tectonic regime, because hydrous phases can form over a wide range of surface temperatures. Geothermal heat flow can likely maintain a sub-ice ocean even at temperatures below the water freezing point \citep[e.g.][]{Warren2002}, and hydrous silicates are stable at temperatures up to $\approx 500-700^{\circ}$ C \citep[e.g.][]{Hacker2003}, meaning these phases could even form under a steam atmosphere.   

However, forming weak faults in the near surface environment is not sufficient for plate tectonics; weakening of the mid- and lower-lithosphere, where strength is maximum, is also necessary. The deep lithosphere is initially dry due to dehydration during mid-ocean ridge melting \citep{hirth1996,evans2005}, and is deeper than hydrothermal circulation can reach \citep{Gregory1981}. Thus, a mechanism capable of hydrating the mid-lithosphere is necessary for water weakening to be viable. One possible mechanism is the ingestion of water along deep thermal cracks \citep{Korenaga2007}. Deep cracking is a result of thermal stresses that arise in the cooling lithosphere. The thermal stresses are proportional to the temperature difference between the cold lithosphere and the mantle interior, so climate can directly influence hydration of the mid-lithosphere. Higher surface temperatures could potentially lead to shallow cracks that leave the mid-lithosphere dry and strong. The surface temperature range where thermal cracking is effective has not been explored, but temperatures would likely have to be on the order of hundreds of degrees hotter than the present day Earth to impede plate tectonics.

An even stronger role for climate stems from the influence of surface temperature on convective stresses. High surface temperatures decrease stresses by lowering the negative thermal buoyancy of the lithosphere (e.g. $\Delta T$ decreases, resulting in lower $\tau_m$ from equations \eqref{tau_m}-\eqref{Ra}). If the convective stress drops below the yield stress, then stagnant lid convection ensues. \cite{Lenardic2008} found that increasing Earth's present day surface temperature by $\sim 100$ K or more is sufficient to induce stagnant lid convection, even with a weak, hydrated lithosphere \citep[see also][]{Weller2015}.

\begin{figure}
\includegraphics[scale = 0.5]{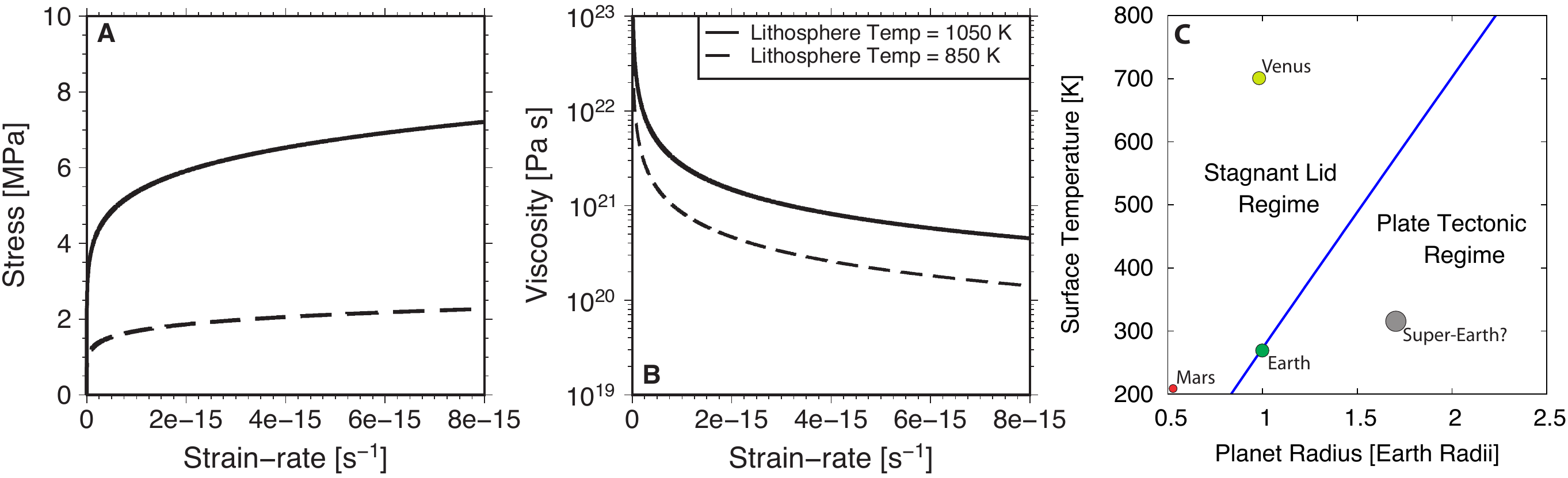}
\caption{\label {fig:damage} Effective stress versus strain-rate relationship (A) and viscosity versus strain-rate relationship (B) resulting from grain-damage (equations \eqref{eqheal}-\eqref{dam_const_law} with $E_h = 500$ kJ mol$^{-1}$, $E_v = 300$ kJ mol$^{-1}$, $m=2$, $p=4$, and $f_A=10^{-4}$). A lower temperature in the mid-lithosphere, which can result from a lower surface temperature, enhances viscous weakening, i.e. lower viscosities result from the same strain-rate.  As a result, the plate tectonic regime is found at low surface temperatures while the stagnant lid regime is found at high temperatures, as shown by the regime diagram from \cite{Foley2012} (C, reprinted with permission). Note that the boundary between the plate-tectonic and stagnant lid regimes in (C) is normalized to the Earth's present day state. The exact position of the boundary between Earth and Venus could not be constrained by the simple models used in \cite{Foley2012}. }  
\end{figure}

When a grainsize reduction mechanism is considered, climate controls whether plate-like convection can occur through the influence of temperature on lithospheric grain-growth rates (Figure \ref{fig:damage}). Grain-growth is faster at higher temperatures (see \eqref{eqheal}), and higher grain-growth rates impede the formation of weak shear zones by acting to increase grainsize (see \eqref{steady_state}). Thus high surface temperatures, which lead to higher temperatures in the mid-lithosphere, promote stagnant lid convection \citep{Landuyt2009a,Foley2012}. \cite{Foley2014_initiation} found that surface temperatures need to reach $\approx 500-600$ K for an Earth-like planet to enter a stagnant lid regime due to enhanced grain-growth. Increasing surface temperature even further would eventually lead to a state where the lithosphere has a low enough viscosity to ``subduct" even without any of the rheological weakening mechanisms discussed in this section (i.e. mantle convection would resemble constant viscosity convection). However, for this style of ``subduction" to occur, the viscosity ratio between the surface and mantle interior must be less than $\approx 10^4$ \citep{slava1995}, which requires a surface temperature of $>1000$ K with an Earth-like mantle temperature of $\approx 1600$ K.  

Summarizing the results from both mechanisms, cold surface temperatures are generally favorable for plate tectonics because they promote grainsize reduction and boost convective stresses. Moreover, cold surface temperatures are unlikely to prevent water weakening because geothermal heat flow can maintain liquid water beneath an ice covered ocean. However, high temperatures, in the range of 400-600 K, can cause stagnant lid convection by dropping convective stresses, increasing grain-growth rates, and possibly suppressing thermal cracking. The influence of climate on tectonic regime is also a leading a hypothesis for the lack of plate tectonics on Venus, as the Venusian surface temperature, 750 K, is easily hot enough to shut down plate tectonics based on the geodynamical models discussed in this section \citep{Lenardic2008,Landuyt2009a}.  

\subsection{Importance of mantle temperature and other factors}
\label{sec:pt_other_factors}

Naturally climate is not the only important factor governing whether terrestrial planets will have plate tectonics.  Mantle temperature, in particular, has been shown to have a major influence on a planet's tectonic regime. Mantle temperature modulates mantle convective stresses and lithospheric grain-growth rates, similar to the influence of surface temperature discussed in \S \ref{sec:climate_tectonics}, and mantle temperature controls the thickness of the oceanic crust formed at ridges, which affects the lithosphere's buoyancy with respect to the underlying mantle. All of these effects act to inhibit plate tectonics when interior temperatures are high, implying that planets with high internal heating rates or young planets, which are hotter because they have had less time to lose their primordial heat \citep[e.g.][]{abbott1994,Korenaga2006,Labrosse2007,Herzberg2010} (see also \S \ref{sec:core_mantle}), will be less likely to have plate tectonics. However, there is disagreement over just how important mantle temperature is in dictating a planet's mode of surface tectonics. 

Increasing mantle temperature drops the mantle viscosity, $\mu_m$. From equations \eqref{tau_m}-\eqref{Ra}, mantle stress scales as $\mu_m^{1/3}$. Thus higher mantle temperatures lead to lower convective stresses, and, if stresses drop below the yield strength, stagnant lid convection in viscoplastic models \citep{ONeill2007b,Moore2013,Stein2013}; this effect may have even caused stagnant lid convection, possibly with episodic subduction events, on the early Earth \citep{ONeill2007b,Moore2013}. The influence of mantle temperature on stress can also lead to hysteresis in viscoplastic mantle convection models \citep{Crowley2012,Weller2012,Weller2015}. Stagnant lid convection results in higher interior temperatures, and thus lower stresses, than mobile lid convection, due to low heat flow across the thick, immobile lithosphere. It is therefore harder to initiate plate tectonics starting from a stagnant lid initial condition than it is to sustain plate tectonics on a planet where it is already in operation.  Such hysteresis would also mean that a planet that loses plate tectonics will be unlikely to restart it at a later time, even before the couplings between climate and the magnetic field are taken into account.  

High mantle temperatures also play a role in plate generation via grainsize reduction, because they result in a warmer mid-lithosphere, and thus higher grain-growth rates. However, the influence of mantle temperature on lithospheric grain-growth rate is weaker than that of surface temperature, so elevated mantle temperatures alone are not capable of shutting down plate tectonics \citep{Foley2014_initiation}, in contrast to viscoplasticity. Furthermore, the weak role of mantle temperature in plate generation means that the hysteresis loops seen in viscoplastic models may be less prominent with a grainsize reduction mechanism, though future study is needed to confirm this.   

Mantle temperature also determines the thickness of oceanic crust generated at spreading centers, with hotter temperatures generally leading to a thicker crust \citep[e.g.][]{White1989}.  As crust is less dense than the underlying mantle, cooling of the lithospheric mantle is necessary for the lithosphere as a whole to become convectively unstable \citep[e.g.][]{Oxburgh1977}.  A very thick crust then requires a long cooling time, such that a thick lithospheric mantle root can form, in order to create a negatively buoyant lithospheric column. Forming a thick lithospheric mantle root is problematic, especially under high mantle temperature conditions, because the lithospheric mantle can delaminate before the lithosphere as a whole reaches convective instability, preventing subduction and plate tectonics \citep[e.g.][]{davies1992b}. The ability of crustal buoyancy to preclude plate tectonics on both the early Earth \citep{davies1992b,Vlaar1985,Vanthienen2004}, and on exoplanets \citep{Kite2009}, has been discussed by many authors. However, the transition from basalt to compositionally dense eclogite, which occurs at shallow depths of $\approx 50$ km on Earth \citep[e.g.][]{Hacker1996}, allows episodic subduction events to occur despite a thick, buoyant crust in models of the early Earth \citep{Vanthienen2004b} and super-Earth exoplanets \citep{ORourke2012}. Moreover, even stable, modern day plate tectonics can potentially still operate because the same melting process that creates the crust also dehydrates and stiffens the sub-crustal mantle; this stiff lithospheric mantle then resists delamination and allows thick, negatively buoyant lithosphere to form through conductive cooling \citep{Korenaga2006}. 

It is also important to point out that other factors, in particularly composition and size, can potentially have a major impact on whether plate tectonics can take place on a rocky planet.  However, both the influence of size and bulk composition on a planet's tectonic regime are not well understood.  There is still significant disagreement over whether increasing planet size makes plate tectonics more or less likely, owing mainly to uncertainties in lithospheric rheology, the pressure dependence of mantle viscosity and thermal conductivity, and the expected radiogenic heating budget of exoplanets (see also \S \ref{sec:sum_exoplanets}). Likewise, little is known about the key material properties for planetary mantles with a significantly different composition than Earth.  Understanding how these factors influence a planet's propensity for plate tectonics and overall evolution is a vital area of future research.  

\section{Climate regulation and the long-term carbon cycle}
\label{sec:climate_reg_cc}

A primary factor determining whether a planet's climate will be conducive to plate tectonics is the degree of greenhouse warming. Venus has $\approx 90$ bar of CO$_2$ in its atmosphere resulting in a surface temperature of $\approx 750$ K that is unfavorable for plate tectonics, while Earth only has $\approx 4 \times 10^{-4}$ bar of atmospheric CO$_2$, and thus has a temperate climate that allows for plate tectonics. However, on Earth there is enough CO$_2$, anywhere from $\approx 60-200$ bar \citep[e.g.][]{Sleep2001b}, locked in carbonate rocks at the surface and in the mantle to cause very hot surface temperatures of $\approx 400-500$ K \citep{Kasting1986}. Thus a key for maintaining plate tectonics on a planet is the ability to prevent large quantities of CO$_2$ from building up in the atmosphere. On planets possessing liquid water, like Earth, this can be accomplished by silicate weathering at the surface and on the seafloor. Crucially, the weathering rate is sensitive to climate, with higher temperatures leading to larger weathering rates (and thus higher CO$_2$ drawdown rates), meaning that weathering acts as a negative feedback mechanism that works to maintain temperate climate conditions \citep{walker1981,berner1983,TajikaMatsui1990,Brady1991,TajikaMatsui1992,Berner1997,Berner2004}. Planets that lack liquid water, like Venus, have no mechanism for regulating atmospheric CO$_2$. However, as we show in this section, simply possessing water does not guarantee that weathering can maintain climates within a range favorable for plate tectonics. A sufficient supply of fresh, weatherable rock at the surface is also needed. We further argue that plate tectonics enhances the supply of fresh rock to the surface, opening the possibility that plate tectonics and the long-term carbon cycle act as a self-sustaining feedback mechanism in some cases. 

\subsection{Modeling the long-term carbon cycle}
\label{sec:weather_feedback}

\begin{figure}
\includegraphics[scale = 0.75]{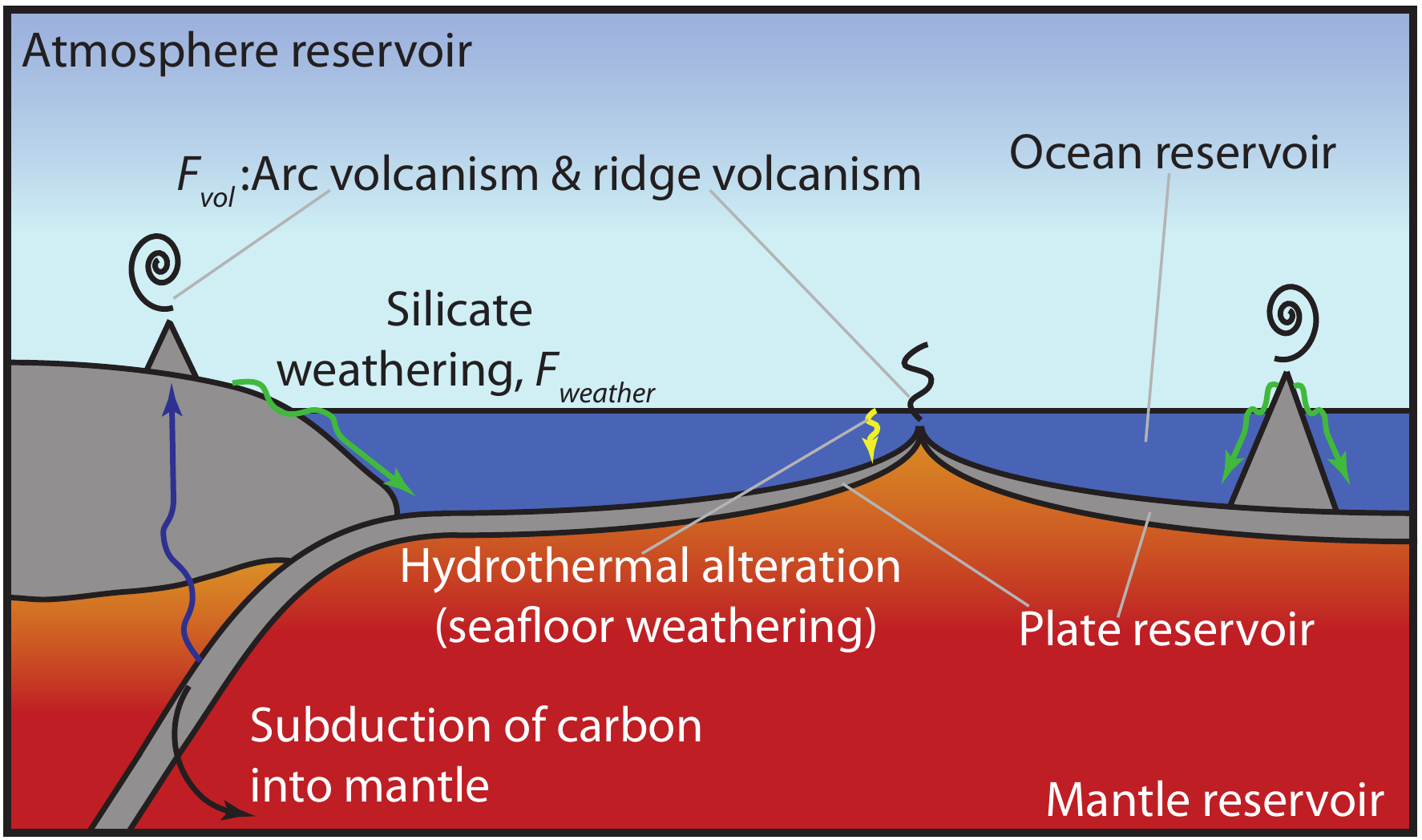}
\caption{\label {fig:c_cycle_model} Schematic diagram of the global carbon cycle after \cite{Foley2015_cc}.}  
\end{figure}

Figure \ref{fig:c_cycle_model} gives a simple illustration of the global carbon cycle where four main reservoirs of carbon are considered (see \cite{Berner2004} and \cite{Ridgwell2005} for detailed reviews): the seafloor, the mantle, the atmosphere, and the ocean (the atmosphere and ocean reservoirs are often combined because equilibration between these two reservoirs is essentially instantaneous on geologic timescales \citep[e.g.][]{Sleep2001b}). Carbon dioxide in the atmosphere is consumed by weathering reactions with silicate rocks, producing calcium, magnesium, and bicarbonate ions that flow from groundwater into rivers, eventually draining into the ocean (see \cite{Gaillardet1999} for a global compilation of CO$_2$ consumption via chemical weathering). Once in the ocean, these ions recombine to precipitate carbonates, leading to an overall net transfer of CO$_2$ from the atmosphere to carbonate rocks on the seafloor. Hydrothermal alteration of basalt can also act as a sink for CO$_2$ dissolved in the oceans, and given rapid equilibration between the atmosphere and ocean, a sink for atmospheric CO$_2$ as well \citep{Staudigel1989,Alt1999,Gillis2011}.  Carbonates on the seafloor, both in the form of sediments and altered basalt, are subducted into the mantle at trenches.  Here, a fraction of the carbon devolatilizes and returns to the atmosphere through arc volcanoes, with the remaining carbon being recycled to the deep mantle.  To close the cycle, mantle carbon is degassed through mid-ocean ridge and plume volcanism back to the atmosphere and ocean reservoirs.

The balance between weathering and volcanic outgassing dictates atmospheric CO$_2$ content. This balance is formulated as
\begin{equation}
\eqlbl{balance}
F_{weather} + F_{sfw} = F_{arc} + F_{ridge},
\end{equation}
where $F_{weather}$ is the silicate weathering flux on land, $F_{sfw}$ is the seafloor weathering flux, $F_{arc}$ is the flux of CO$_2$ degassing from volcanic arcs, and $F_{ridge}$ is the degassing flux at ridges. Balance between silicate weathering and degassing typically occurs rapidly, on a timescale of $ \sim 1$ Myr or less \citep[e.g.][]{Sundquist1991,Berner1997,Driscoll2013,Foley2015_cc}, so assuming that weathering always balances degassing is reasonable when studying long-term climate evolution. The arc degassing flux is typically written as
\begin{equation}
\eqlbl{arc}
F_{arc} = \frac{f v_p L R_p}{A_p} ,
\end{equation} 
where $f$ is the fraction of subducted carbon that degasses, $v_p$ is the plate speed, $L$ is the length of subduction zone trenches, $R_p$ is the amount of carbon on the seafloor, and $A_p$ is the area of the seafloor. The ridge degassing flux is given by 
\begin{equation}
F_{ridge} = \frac{f_d v_p L d_{melt} R_{man}}{V_{man}} ,
\end{equation}
where $f_d$ is the fraction of upwelling mantle that degasses, $L$ is the length of ridges (assumed equal to the length of trenches), $d_{melt}$ is the depth where mid-ocean ridge melting begins, $R_{man}$ is the amount of carbon in the mantle, and $V_{man}$ is the volume of the mantle \citep[e.g.][]{TajikaMatsui1990,TajikaMatsui1992,Sleep2001b,Driscoll2013,Foley2015_cc}.

Silicate weathering can exert a negative feedback on climate because mineral dissolution rates, and hence CO$_2$ drawdown rates, increase with increasing temperature and precipitation \citep[e.g.][]{walker1981,berner1983} (see \cite{Kump2000} and \cite{Brantley2014} for reviews on mineral dissolution kinetics and atmospheric CO$_2$). A similar link between climate and weathering rate can be seen in many field studies \citep[e.g.][]{Velbel1993,White_Blum1995,Dessert2003,Gislason2009,White2014,Viers2014}. However, many other locations show no link between weathering and climate; instead the weathering flux is linearly related to the physical erosion rate \citep{Stallard1983,Edmond1995,Gaillardet1999,Oliva1999,Millot2002,Dupre2003,Riebe2004,West2005,Viers2014}. These conflicting field observations are likely due to the distinction between ``kinetically limited" (or ``reaction limited") weathering and ``supply limited" weathering. When weathering is kinetically limited, the kinetics of mineral dissolution controls the weathering rate. However, when weathering is supply limited, the supply of fresh rock brought to the weathering zone by erosion controls the weathering rate.

Supply limited weathering occurs when the reaction between silicate minerals and CO$_2$ runs to completion in the regolith, requiring physical erosion to expose fresh bedrock for weathering to continue. The supply limited weathering flux ($F_{w_s}$) can be written as \citep{Riebe2004,West2005,Mills2011,Foley2015_cc}
\begin{equation}
F_{w_s} = \frac{A_{land} E \chi \rho_{cc}}{\bar{m}} ,
\end{equation}
where $A_{land}$ is the surface area of all exposed land, $E$ is the physical erosion rate, $\chi$ is the fraction of reactable elements in the crust, $\rho_{cc}$ is the density of the crust, and $\bar{m}$ is the molar mass of reactable elements. The kinetically limited weathering flux, $F_{w_k}$, is sensitive to climate and can be written as \citep[e.g.][]{walker1981,berner1983,TajikaMatsui1990,TajikaMatsui1992,Berner1994,Sleep2001b,Berner2004,Driscoll2013}
\begin{equation}
\eqlbl{f_weather}
F_{w_k} = F_w^* \exp{\left [ \frac{E_a}{R_g} \left(\frac{1}{T^*} - \frac{1}{T} \right) \right ]} \left(\frac{P_{CO_2}}{P^*_{CO_2}}\right)^{\beta} \left(\frac{R}{R^*} \right)^{\alpha} \left(\frac{f_{land}}{f_{land}^*} \right) ,
\end{equation}
where $F^*_w$ is the present day rate of atmospheric CO$_2$ drawdown by silicate weathering ($F^*_w \approx 6 \times 10^{12}$ mol yr$^{-1}$, or half the estimate of \cite{Gaillardet1999} because half of the CO$_2$ removed by weathering is re-released to the atmosphere when carbonates form \citep{Berner2004}), $E_a$ is the activation energy for mineral dissolution, $R_g$ is the universal gas constant, $T$ is temperature, $P_{CO_2}$ is the partial pressure of atmospheric CO$_2$, $R$ is the runoff, $f_{land}$ is the land fraction (subaerial land area divided by Earth's surface area), and $\beta$ and $\alpha$ are constants. Stars represent present day values.  

The exponential term in \eqref{f_weather} describes the temperature sensitivity of mineral dissolution rates, where typical values of $E_a$ range from $\approx 40-50$ kJ mol$^{-1}$ \citep[e.g.][]{Brady1991}. The $P_{CO_2}$ term describes the direct dependence of silicate weathering rates on atmospheric CO$_2$ concentration, which results from the role of pH in mineral dissolution.  Specifically, more acidic pH values cause faster dissolution rates. However, when dissolution is caused by organic acids produced by biology (mainly plants on the modern Earth), soil pH is fixed, and the direct dependence of weathering rate on $P_{CO_2}$ is weak (i.e. $\beta$ in \eqref{f_weather} is approximately 0.1 or less \citep{Volk1987,Berner1994,Sleep2001b}). Without biologically produced acids, carbonic acid formed when atmospheric CO$_2$ dissolves in rainwater sets the soil pH, making the direct dependence of weathering rate on $P_{CO_2}$ stronger ($\beta \approx 0.5$ \citep{Berner1992}). The runoff term adds an additional climate feedback, as precipitation rates increase with temperature ($\alpha \approx 0.6-0.8$ \citep{Berner1994,Bluth1994,West2005}; see \cite{Berner1994} for a detailed discussion of how silicate weathering scales with runoff). Finally, the weathering rate scales with land fraction, because decreasing the total area of land undergoing weathering lowers the total weathering flux. 

The supply limited and kinetically limited weathering fluxes can be combined into a total weathering flux following \cite{Gabet2009}, \cite{Hilley2010}, and \cite{West2012} (see \cite{Foley2015_cc} for a full derivation) as
\begin{equation}
F_{weather} =  F_{w_s} \left [ 1 - \exp{\left (-\frac{F_{w_k}}{F_{w_s}} \right)} \right ].
\end{equation}
When $F_{w_k}$ is below the supply limit, $F_{w_s}$, the overall weathering rate is approximately equal to the kinetically limited weathering rate, i.e. $F_{weather} \approx F_{w_k}$. However, when $F_{w_k}$ reaches or exceeds the supply limit, the overall weathering rate is fixed at $F_{w_s}$; i.e. $F_{weather}$ can no longer increase with increasing surface temperature or atmospheric CO$_2$ content once the supply limit to weathering has been reached. The present day Earth consists of both regions undergoing supply limited weathering and those undergoing kinetically limited weathering. However, kinetically limited weathering appears to be the dominant component of the global silicate weathering flux \citep{West2005}. 

Finally, CO$_2$ drawdown also occurs via seafloor weathering on mid-ocean ridge flanks \citep{Alt1999}. Seafloor weathering is thought to be a smaller present day CO$_2$ sink than weathering on land, and whether it exerts a strong feedback on climate is debated \citep{Caldeira1995,Brady1997,Berner2004,Coogan2013,Coogan2015}. Many studies have assumed a weak dependence on $P_{CO_2}$, based on the experiments of \cite{Brady1997}, and formulate the seafloor weathering flux as \citep[e.g.][]{Sleep2001b,Mills2014,Foley2015_cc}
\begin{equation}
\eqlbl{sfw}
F_{sfw} = F_{sfw}^* \left(\frac{v_p}{v_p^*}\right) \left(\frac{P_{CO_2}}{P_{CO_2}^*} \right)^{\xi} ,
\end{equation} 
where the present day seafloor weathering flux $F^*_{sfw} \approx 1.75 \times 10^{12}$ mol yr$^{-1}$ \citep{Mills2014} and $\xi \approx 0.25$. However, both a stronger $P_{CO_2}$ dependence or a direct temperature feedback are possible. Seafloor weathering can also become ``supply limited," because only a finite amount of CO$_2$ can be stored in the ocean crust. \cite{Sleep2001a} and \cite{Sleep2014} estimate that only the top $\approx 500$ m of ocean crust can be completely reacted with CO$_2$, meaning no more than $\approx 4 \times 10^{21}$ mol (or $\approx 30$ bar), $\approx$ 1/10-1/2 Earth's total CO$_2$ budget, can be locked in the seafloor at any one time. 

\subsection{The silicate weathering climate feedback and plate tectonics}

\begin{figure}
\includegraphics[scale = 0.5]{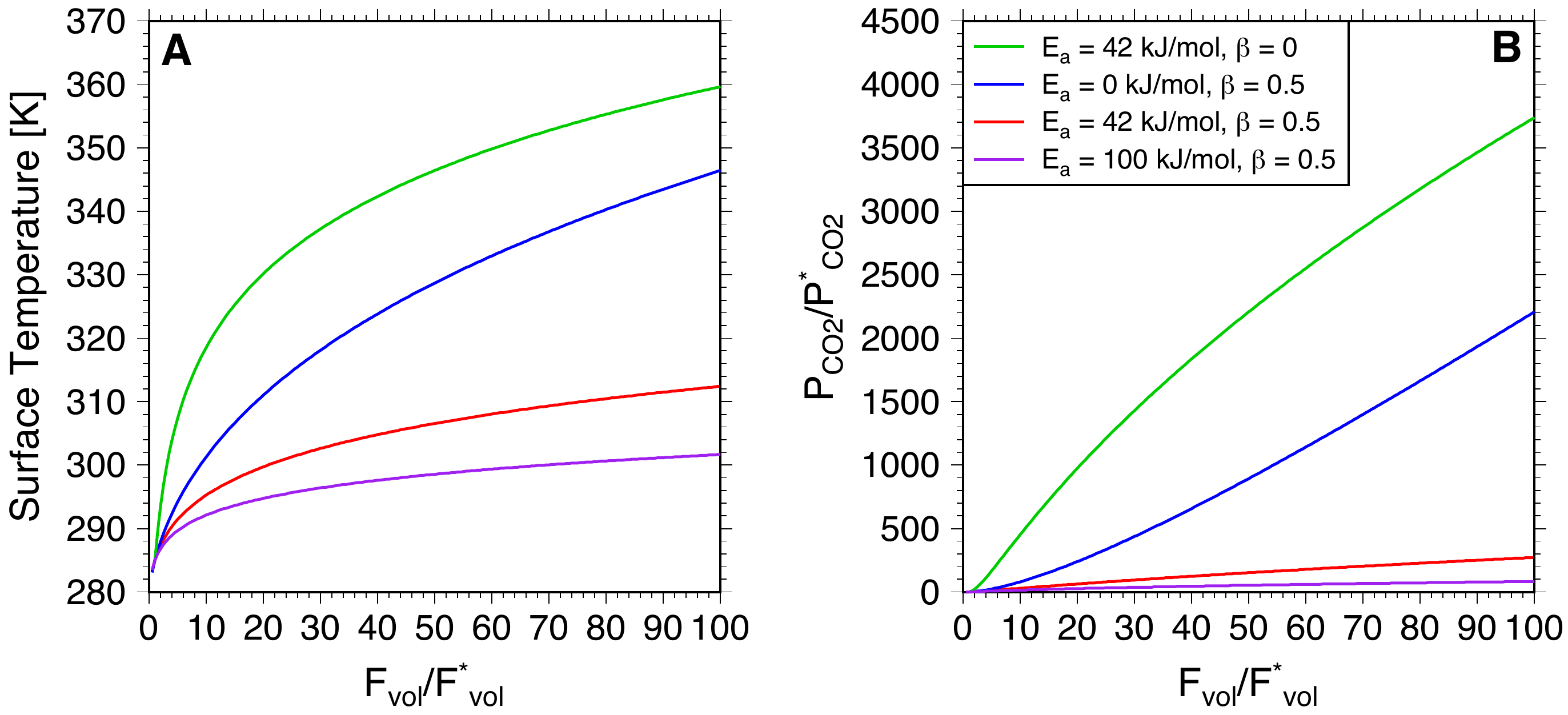}
\caption{\label {fig:fvol} Response of surface temperature (A) and partial pressure of atmospheric CO$_2$, normalized by the present day value (B), to increases in the CO$_2$ degassing rate (where $F_{vol}/F^*_{vol}$ is the volcanic outgassing rate normalized by the present day value). Different values of $E_a$ and $\beta$ are used, as described in the legend, to show how stronger climate feedbacks cause climate to be less sensitive to degassing rate. Typical values for the Earth are $E_a = 42$ kJ mol$^{-1}$ and $\beta = 0.5$ (red line). }  
\end{figure}

Kinetically limited weathering is able to prevent massive quantities of CO$_2$ from building up in the atmosphere, because the weathering rate increases with increasing atmospheric CO$_2$ content. Considering a balance between a given degassing flux, $F_{vol} = F_{arc} + F_{ridge}$, and weathering, the atmospheric CO$_2$ content as a function of degassing rate can be calculated (Figure \ref{fig:fvol}). When the degassing rate is increased, only a small increase in atmospheric CO$_2$ concentration is needed to bring weathering back into balance with degassing, because the weathering rate is a strong function of $P_{CO_2}$. The stronger the dependence of weathering on atmospheric CO$_2$, the less sensitive $P_{CO_2}$ is to variations in the degassing rate. It is this ability to balance the volcanic outgassing flux with minimal changes in $P_{CO_2}$ that allows weathering to prevent extremely hot climates, and thus maintain conditions favorable for plate tectonics. 

The negative feedback between climate and weathering is also important for habitability, because it acts to stabilize climate in response to variations in solar luminosity.  Stars increase in luminosity as they age; for example the sun's luminosity was 70 \% its present day level during the Archean \citep{Gough1981}. The weathering feedback acts to partially counteract this change in luminosity by allowing higher CO$_2$ levels when luminosity is low and lower CO$_2$ levels when luminosity is high.  However, this topic has been extensively studied \citep[e.g.][]{walker1981,Franck1999,Sleep2001b,Abbot2012}, and going into more detail is beyond our scope.  

\begin{figure}
\includegraphics[scale = 0.5]{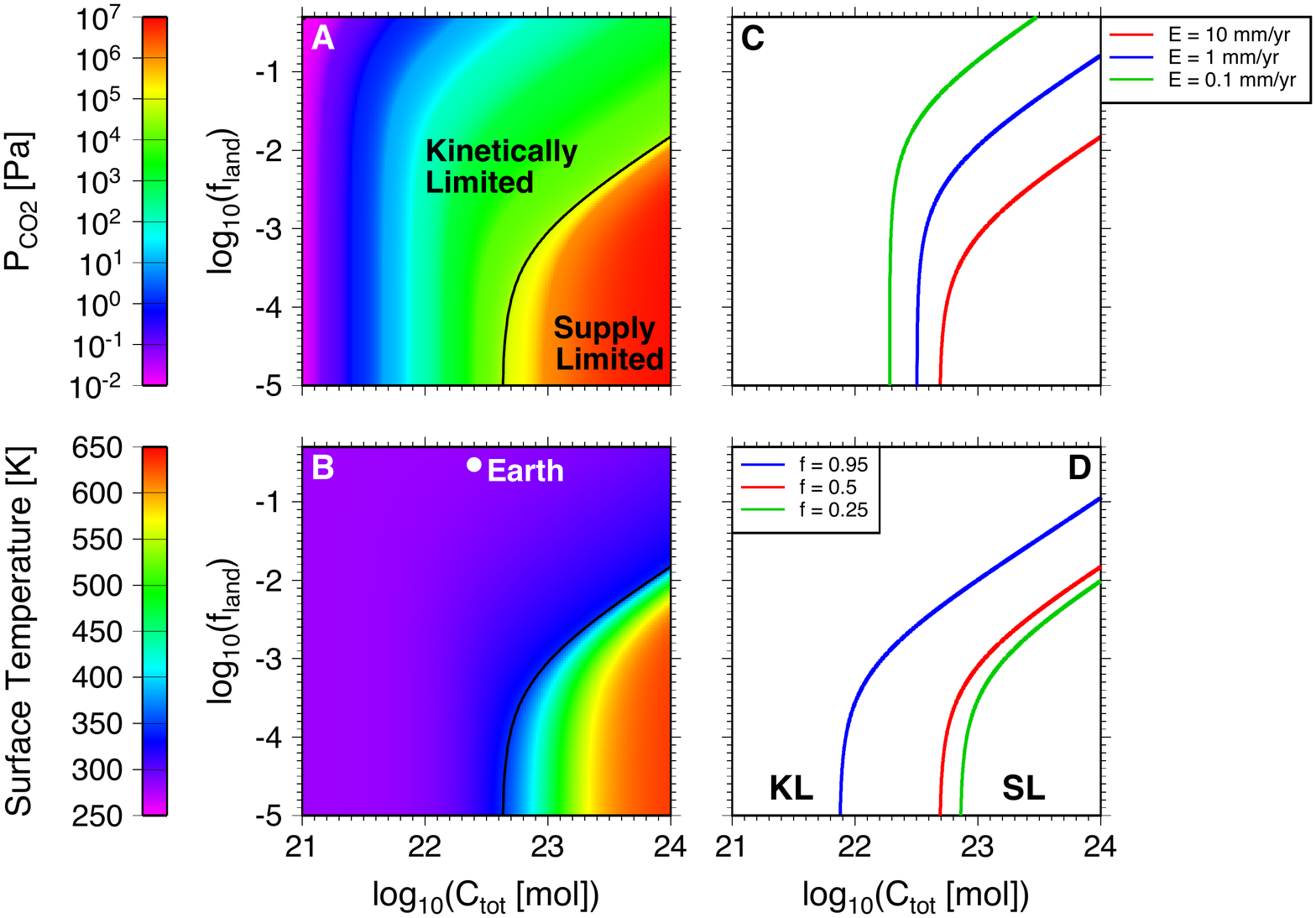}
\caption{\label {fig:supply_lim} Partial pressure of atmospheric CO$_2$ (A) and surface temperature (B) as a function of total planetary CO$_2$ budget and land fraction ($f_{land}$) after \cite{Foley2015_cc}. Kinetically limited weathering occurs at large land fractions and low planetary CO$_2$ budgets, resulting in temperate surface temperatures, while supply limited weathering results in extremely hot climates when land fraction is low or total CO$_2$ budget is high.  The estimated location of the Earth in land fraction-planetary CO$_2$ budget space is shown in panel (B).  The boundary between the kinetically limited and supply limited weathering regimes is shown in panel (C) as a function of erosion rate, $E$, and in panel (D) as a function of the fraction of subducted carbon that degasses at arcs, $f$ (where the label KL refers to the kinetically limited regime and SL refers to the supply limited regime). }  
\end{figure}

On the other hand, when weathering becomes supply limited, it can no longer increase with atmospheric CO$_2$ level and is therefore unable to balance the degassing flux (globally supply limited weathering requires that $F_{vol} \geq F_{w_s}$, otherwise weathering would not be supply limited).  As a result, atmospheric CO$_2$ accumulation from volcanic outgassing continues unabated until the mantle and plate reservoirs are depleted in carbon, and extremely hot climates, that are unfavorable for plate tectonics, prevail (Figure \ref{fig:supply_lim}). Small land areas or low erosion rates lead to supply limited weathering, because both factors limit the supply of fresh rock to the surface, and hence lower $F_{w_s}$ \citep[see][for details]{Foley2015_cc}. Alternatively, factors that increase the degassing rate can also drive a planet into the supply limited weathering regime, even with a large land area or high erosion rates. Planets with large total CO$_2$ inventories are more susceptible to supply limited weathering because degassing rates are higher. Another important factor is the fraction of subducted carbon that reaches the deep mantle, instead of devolatilizing and returning to the atmosphere at arcs ($f$ from equation \eqref{arc}). When more carbon can be subducted and stored in the mantle, $F_{arc}$ is lower and kinetically limited weathering is easier to maintain (Figure \ref{fig:supply_lim}D). The fraction of subducted carbon that degasses at arcs is not well constrained, and the physical and chemical processes controlling this number are poorly understood \citep[e.g.][]{Kerrick2001,Dasgupta2010,Ague2014,Kelemen2015}. Moreover, $f$ is likely a function of mantle temperature, as more slab CO$_2$ will devolatilize during subduction into a hotter mantle \citep[e.g.][]{Dasgupta2010}. Thus planets with hot mantles, as expected for young planets or those with high radiogenic heating budgets (see \S \ref{sec:core_mantle}), may be more susceptible to supply limited weathering. Better constraints on carbon subduction and devolatilization are clearly needed for understanding global climate feedbacks. 

{Figure \ref{fig:supply_lim} assumes that seafloor weathering follows \eqref{sfw}, as in \cite{Foley2015_cc}, which provides only a modest climate feedback. However, including a stronger climate feedback does not prevent extremely hot climates from forming when weathering on land is supply limited, because} seafloor weathering will also become supply limited when all of the accessible basalt is completely altered \citep{Sleep2001a,Sleep2014}. With the ocean crust unable to take in any more carbon, a CO$_2$ rich atmosphere still forms \citep{Foley2015_cc}. Furthermore, the high surface temperatures predicted for the supply limited weathering regime could potentially lead to rapid water loss to space or even trigger a runaway greenhouse. Such scenarios are not explicitly modeled here, but would reinforce the fact that supply limited weathering causes a climate state that is unfavorable for plate tectonics; losing water would prevent volatile weakening of the lithosphere, and a runaway greenhouse climate results in temperatures even higher than those shown in Figure \ref{fig:supply_lim}.  

\subsection{Extent of coupling between climate and surface tectonics}
\label{sec:pt_cc}

Plate tectonics helps to sustain kinetically limited weathering, and therefore prevent extremely hot climates from forming, by acting to maintain high erosion rates and large subaerial land areas through orogeny and volcanism. Erosion rates are highest in rapidly uplifting, tectonically active areas \citep[e.g.][]{Portenga2011}, and orogenic processes are the primary cause of such uplift. Without tectonic uplift, erosion rates would decay to very low values because the surface can only erode as quickly as uplift creates topography. Orogeny results from plate tectonic processes, such as continent-continent collisions, island arc formation, and accretion of arcs to cratons, so plate tectonics is essential for keeping erosion rates on Earth high. Tectonic uplift and physical erosion have long been thought to be important for weathering and climate \citep[e.g.][]{Raymo1992,Maher2014}, and the transition between kinetically limited and supply limited weathering provides an exciting new mechanism for explaining this influence \citep[e.g.][]{Kump1997,Froelich2014}. 

Volcanism also creates topography and supplies fresh rock to the surface, so the fact that plate tectonics causes widespread subaerial volcanism (e.g. through arc, plume, and other hot spot melting), is important for sustaining high erosion rates as well. Perhaps the more important role for volcanism, however, is in creating exposed land, as large land areas also increase the supply of weatherable rock. In particular, continental crust, which makes up the large majority of Earth's exposed land, is thought to be predominantly created by subduction zone volcanism \citep[e.g.][]{Rudnick1995,Cawood2013}, a process that is unique to plate tectonics. Though hotspot or plume volcanism, the dominant mode of volcanism for a stagnant lid planet, can still create land and continental crust \citep[e.g.][]{Stein1996,Smithies2005}, plate tectonics clearly enhances subaerial land.

In fact, on a planet with a large area of exposed land, like Earth, the high erosion rates provided by plate tectonics may not be necessary for maintaining kinetically limited weathering. On Earth, typical erosion rates for flat-lying, tectonically inactive regions are on the order of 0.01 mm yr$^{-1}$ \citep{Portenga2011,Willenbring2013}, which gives a supply-limited weathering flux on the order of $10^{13}$ mol yr$^{-1}$ when combined with Earth's large land area of $\approx 1.5 \times 10^{14}$ m$^2$. However, as half of the CO$_2$ removed by silicate weathering is re-released to the atmosphere when carbonates form \citep[e.g.][]{berner1983}, the net rate of CO$_2$ drawdown is $\sim 5 \times 10^{12}$ mol yr$^{-1}$. The present day degassing flux is $\approx (6-10) \times 10^{12}$ mol yr$^{-1}$ \citep[e.g.][]{Marty1998,Sleep2001b,Burton2013}, which would be significantly lower if Earth were in a stagnant lid regime. \cite{Marty1998} estimate $\approx 3 \times 10^{12}$ mol yr$^{-1}$ of CO$_2$ degassing from plumes, which is a reasonable approximation for Earth's degassing flux were it in a stagnant lid regime. As the stagnant lid degassing flux is lower than the supply-limited weathering flux, kinetically limited weathering is possible on a hypothetically stagnant lid Earth. 

Therefore plate tectonics and the long-term carbon cycle probably only act as a self-sustaining feedback on planets with small land areas or large planetary CO$_2$ inventories, where high erosion rates are needed to prevent supply limited weathering (i.e. planets that would lie near the boundary between kinetically limited and supply limited weathering in Figure \ref{fig:supply_lim}). Unfortunately a more quantitative estimate is not possible with our current level of understanding on how tectonics modulates erosion and weathering. Nevertheless, given that volatile acquisition during terrestrial planet formation can vary significantly \citep[e.g.][]{Raymond2004}, volatile rich planets where plate tectonics and the carbon cycle are so tightly coupled may be common. Furthermore, even if a planet's exposed land area is large enough to sustain kinetically limited weathering at low erosion rates, plate tectonics may still be important. Plate tectonics may be responsible for creating most of the exposed land through continental crust formation, without which much higher erosion rates would be needed for kinetically limited weathering. Earth may even be an example of such a planet.     

Another aspect of the coupling between plate tectonics and climate is that plate tectonics leads to long-lived CO$_2$ degassing by recycling carbon into the mantle at subduction zones.  This process is important for habitability, because when CO$_2$ degassing rates are low, or cease entirely, snowball climates can form \citep{Kadoya2014}.  However, snowball climates are unlikely to also shut down plate tectonics, so the role of plate tectonics in sustaining CO$_2$ degassing probably does not represent a self-sustaining feedback between tectonics and climate.

The different evolutionary scenarios that result from coupling between climate and surface tectonics are described in \S \ref{sec:summary}. In particular planets at different orbital distances can undergo different evolutions if one planet lies inward of the habitable zone's inner edge and thus lacks silicate weathering. Likewise different initial climate conditions can cause divergent evolutions if an initially hot climate prevents plate tectonics and kinetically limited weathering, or even weathering at all, from transpiring. 

\section{Generation of the magnetic field and its role in atmospheric escape}
\label{sec:whole_planet}

We have described the interactions between the surface environment (atmosphere plus ocean) and the mantle in sections \ref{sec:plate_generation} \& \ref{sec:climate_reg_cc}. In this section we demonstrate that additional interactions exists between the mantle and core, and the geomagnetic field (which is generated by the core dynamo) and atmosphere. The mantle controls the rate at which the core cools, thereby playing a crucial role in maintaining the energy flow necessary to drive convection and dynamo action in the core.  The geomagnetic field provides a shield that holds the solar wind far above the surface (presently at about 9 Earth-radii) so that most high energy particles are diverted and prevented from disrupting the near surface environment.  As a result, magnetic fields may limit the atmospheric escape rate under certain conditions. The magnetic field's influence on escape rate can then have an important control on long-term climate evolution, opening the possibility for an indirect influence of the core dynamo on mantle convection, in addition to the direct role mantle convection plays in driving the dynamo.

\subsection{Core dynamo}  
\label{sec:dynamo}

Earth has maintained an internally generated planetary magnetic field through convective dynamo action in its core over much of its history \citep{tarduno2015}. The presence of a long-lived magnetic field is another unique feature of our planet, as only Mercury and Ganymede also have active magnetic fields today among the terrestrial planets and moons of the solar system \citep{schubert2011}. Internally generated magnetic fields are created by convective dynamo action in a large rotating volume of electrically conductive liquid.  In the terrestrial planets this typically occurs in an iron core. The most commonly cited process for driving dynamo action, and thus the focus of this section, is thermal or compositional convection. However, other driving mechanisms, such as gravitational tides \citep[e.g.][]{cebron2014} or precipitation of Mg initially dissolved in the liquid iron \citep{orourke2016} are possible.  

Thermal convection occurs when the core heat flow, $Q_{cmb}$, exceeds the heat conducted adiabatically through the core.  Recent estimates place the core's conductive heat flow at 13-15 TW \citep{Driscoll2014,pozzo2014}. A high core heat flow is also inferred from revisions to the dynamics of mantle plumes, which are thought to be generated at the core-mantle boundary (CMB) and be indicative of the CMB heat flow \citep{bunge2005,zhong2006}.
Even if the total core heat flow is less than that needed to keep the core adiabatically well mixed, compositional buoyancy may still be able to drive convection.  Compositional convection, which can be driven a number of ways, usually involves a phase change in the liquid where a density gradient develops or by the dissolution of an incompatible element due to a change in the solubility.  For example, as Earth's core cools the inner core crystallizes from below, releasing buoyant light element rich fluid into the surrounding iron-rich fluid and driving compositional convection. In this example core heat flow is still important for driving compositional convection, because core cooling is necessary for inner core growth. 

The heat transferred across the core-mantle boundary, $Q_{cmb}$, is controlled by the temperature gradient in the viscous mantle above:
\begin{equation}
Q_{cmb}=A_{cmb}k_{LM}\frac{\Delta T_{LM}}{\delta_{LM}} 
\label{q_cmb} \end{equation}
where $A_{cmb}$ is CMB area, and $k_{LM}$, $\Delta T_{LM}$, and $\delta_{LM}$ are the thermal conductivity, temperature drop, and thickness of the lower mantle thermal boundary layer.  The temperature drop, $\Delta T_{LM}=T_{cmb}-T_{m}$, measures the thermal disequilibrium between the mantle and core.  If the mantle cools efficiently then $\Delta T_{LM}$ is large, resulting in a high $Q_{cmb}$.  The thermal boundary layer thickness, $\delta_{LM}$, can be derived by assuming the boundary layer Rayleigh number (i.e. $\rho g \alpha \Delta T_{LM} \delta_{LM}^3/(\kappa \mu_{LM})$, where $\mu_{LM}$ is the effective viscosity of the lower mantle boundary layer) is at the critical value for convection.   Compositional differences between the lower mantle and the bulk mantle \citep[e.g.][]{sleep1988} can influence the thickness of $\delta_{LM}$; this influence can be incorporated by modifying $\mu_{LM}$ \citep{Driscoll2014} or the critical Rayleigh number \citep{stevenson1983}.  Calculating $\delta_{LM}$ from the boundary layer Rayleigh number implies that a hotter mantle produces a thinner boundary layer, increasing $Q_{cmb}$.  Therefore, the ratio $\Delta T_{LM}/\delta_{LM}$ in (\ref{q_cmb}) implies that the core and mantle will evolve towards thermal equilibrium faster in a hotter mantle and that efficient mantle cooling leads to efficient core cooling.

However, cooling the core too fast can be detrimental to dynamo action in two ways: (1) rapid cooling can lead to complete solidification of the core, preventing fluid motion and dynamo action, and (2) rapid cooling can bring the core into thermal equilibrium with the mantle, which will eventually decrease the cooling rate below the threshold for driving convection.  The later is the eventual fate of every rocky body, but the longer a planet can avoid thermal equilibrium and maintain moderate cooling the longer it will generate a magnetic field.

\subsection{Coupling between core dynamo and tectonic mode}
\label{sec:core_mantle}

As the core cooling rate is determined by the mantle, the surface tectonic mode, which determines the mantle cooling rate, dramatically influences the dynamo.  A mobile lid can accommodate a larger surface heat flow by exposing the mantle's top thermal boundary layer to the surface, while a stagnant lid inhibits cooling by insulating the mantle with a thick conductive layer.


%

To model the cooling and convective power available to drive dynamo action over time, the energy balance of the core and mantle are solved simultaneously.  
Volume averaged temperature evolution can be derived using the secular cooling equation $Q_{i}=-c_i M_i \dot{T}_i$, where $c$ is specific heat and $i$ refers to either mantle ($i = m$) or core ($i = c$) in the mantle and core energy balances \citep[e.g.][]{Driscoll2014}.  Solving for $\dot{T}_m$ and $\dot{T}_c$ in terms of sources and sinks gives the mantle and core thermal evolution equations,
\begin{equation}
 \dot{T}_m=\frac{\left( Q_{cmb} + Q_{rad} -Q_{conv}-Q_{melt} \right)}{M_m c_m  }
\label{dot_T_m} \end{equation}
\begin{equation}
\dot{T}_c= -\frac{ (Q_{cmb}-Q_{rad,c})} {M_c c_c - A_{ic} \rho_{ic} \epsilon_c \frac{d R_{ic}}{d T_{cmb}} (L_{Fe}+E_G) }
\label{dot_T_c} \end{equation}
where $Q_{conv}$ is heat conducted through the lithospheric thermal boundary layer by mantle convection (in W), $Q_{melt}$ is heat loss by mantle melt eruption, and $Q_{rad}$ and $Q_{rad,c}$ are radiogenic heat production in the mantle and core, respectively.  Crustal heat sources are excluded because they do not contribute to the mantle heat budget.  
The denominator of (\ref{dot_T_c}) is the sum of core specific heat and heat released by inner core growth, where $A_{ic}$ is inner core surface area, $\rho_{ic}$ is inner core density, $\epsilon_c$ is a constant that relates average core temperature to CMB temperature, $dR_{ic}/dT_{cmb}$ is the rate of inner core growth as a function of CMB temperature, and $L_{Fe}$ and $E_G$ are the latent and gravitational energy released at the ICB per unit mass. Detailed expressions for heat flows and temperature profiles as functions of mantle and core properties are given in \citet{Driscoll2014} and \citet{driscoll2015b}.

Heat is lost from the mantle via conduction through the upper mantle thermal boundary layer ($Q_{conv}$) and by melt eruption ($Q_{melt}$).  The rate at which this heat is lost is a complex function of mantle temperature, material properties, and style of convection.
A mobile lid implies that the upper mantle thermal boundary layer reaches the planetary surface, so that the convective heat loss is controlled by conduction through this thin boundary layer.  Assuming the boundary layer is at the critical Rayleigh number for convection, the convective heat flow is
\begin{equation}
Q_{conv}=a_m \nu_m^{-\beta}T_m^{\beta+1}  \label{q_conv} ,
\end{equation}
where $\nu_m = \mu_m/\rho$ is temperature-dependent mantle kinematic viscosity and $a_m=8.4\times10^{10}$ W(m$^2$s$^{-1}$K$^{-4}$)$^{1/3}$ and $\beta=1/3$ are constants \citep[see][equation 43]{Driscoll2014}.
In steady state, a stagnant lid has a thicker conductive boundary layer, which can be modeled by decreasing $a_m$ by a factor of $\sim25$ \citep{slava1995}. 

Mobile lid convection therefore favors dynamo action because it efficiently cools the mantle and thus boosts the core heat flow. Stagnant lid convection, on the other hand, can impede core cooling; stagnant lid convection on Venus is a leading hypothesis for why Venus lacks a magnetic field \citep{nimmo2002}. However, stagnant lid convection does not always prevent a core dynamo, as thermal history models commonly show a relatively short period of rapid cooling, regardless of the surface tectonic mode, during which core cooling rates are high and dynamo action is possible. This period of rapid cooling, or thermal adjustment period, results from initially hot interior temperatures that are a consequence of planetary accretion. The thermal adjustment period typically lasts for the first $1-2$ Gyr as the internal temperatures and heat flows adjust to boundary conditions and heat sources. Similar early thermal dynamos have been proposed for the Moon \citep{stegman2003} and Mars \citep{nimmo2000}.  
 
\begin{figure}[ht!]
\includegraphics[width=\linewidth]{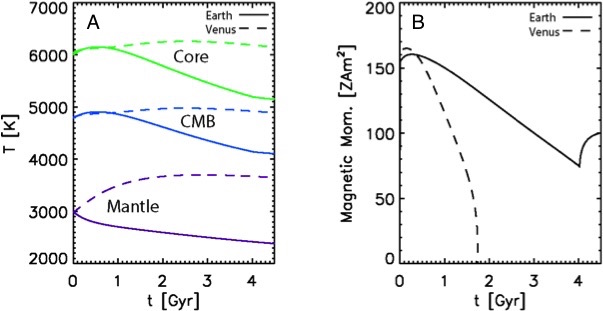}
\caption{Comparison of the predicted thermal (A) and magnetic (B) histories of Earth (solid) and Venus (dashed).  The only difference between the models is the Earth model uses a mobile lid heat flow scaling while the Venus model uses a stagnant lid scaling (see (\ref{q_cmb})-(\ref{q_conv}); \citet{Driscoll2014} contains additional details).}
\label{fig:EV}
\end{figure}

Figure \ref{fig:EV} shows example thermal histories for Earth and Venus, where the Earth model uses a mobile lid heat flow scaling while the Venus model uses a stagnant lid scaling \citep[see][]{Driscoll2014}. The initial mid-mantle and CMB temperatures are assumed to be at the silicate liquidus, which is a reasonable starting temperature following the last large accretion event.
Both models experience an initial thermal adjustment for $1-2$ Gyr, but diverge soon after.  After the initial adjustment the Earth model cools monotonically, maintaining a thermal dynamo and later a thermo-chemical dynamo after inner core nucleation around 4 Gyr, which can be seen as a jump in the predicted magnetic moment (Figure \ref{fig:EV}B).  The core of the Venus model cools slightly during the thermal adjustment period, driving a transient thermal dynamo, but is slowly heated as the mantle and core heat up due to radiogenic heat trapped beneath the stagnant lid.  In this case the core is too hot to solidify, precluding a compositional dynamo.  These models predict Venus maintained a magnetic field for about 1.5-4 Gyr \citep{Driscoll2014}.
During the thermal adjustment period the core and mantle are still strongly coupled, but the cooling rate is less sensitive to the style of mantle convection. Thus the primary role of plate tectonics is to extend the life of the dynamo beyond this thermal adjustment period. 

In addition to normal convective cooling, volcanic heat loss can potentially extend core cooling beyond the thermal adjustment period on a stagnant lid planet, provided most of the melt can reach the surface of the planet and cool \citep{morgan1983,nakagawa2012,armann2012,Driscoll2014}.  Obviously, magmatic heat loss was not efficient enough for Venus, Mars, or the Moon to sustain dynamos, so we are left with the tentative conclusion that a stagnant lid reduces both the long-term convective and volcanic mantle heat loss.

The role of mantle dynamics in dictating magnetic field strength also indirectly links the core to climate. Climate influences the tectonic regime of a planet, and the tectonic regime dictates the core cooling rate through time and thus whether a long-lived magnetic field can be maintained. Therefore a cool climate, being favorable for plate tectonics, will also be favorable for long-term magnetic field generation, and hot climates, to the extent that they lead to stagnant lid convection, will be unfavorable for the dynamo. An important question is then whether the core can exert any influence on climate through the role the magnetic field plays in limiting atmospheric escape. We address this topic next.

\subsection{Magnetic limited escape}
\label{sec:mag_limited_escape}

In describing the escape of a planetary atmosphere it is helpful to think of the limiting escape mechanism rather than the specific physical escape process, as several escape processes typically occur concurrently with a single limiting bottleneck.
The escape limit is typically characterized as being in either the diffusion or energy limited regime, with no connection to the presence of a magnetic field.  
However, it is commonly argued that magnetic field strength should play an important role in atmospheric escape \citep{michel1971,chassefiere1997,lundin2007,dehant2007,lammer2012,owen2014,brain2014}. Below we speculate about how a magnetic field could influence escape. 

\begin{figure}[ht!]
\includegraphics[width=\linewidth]{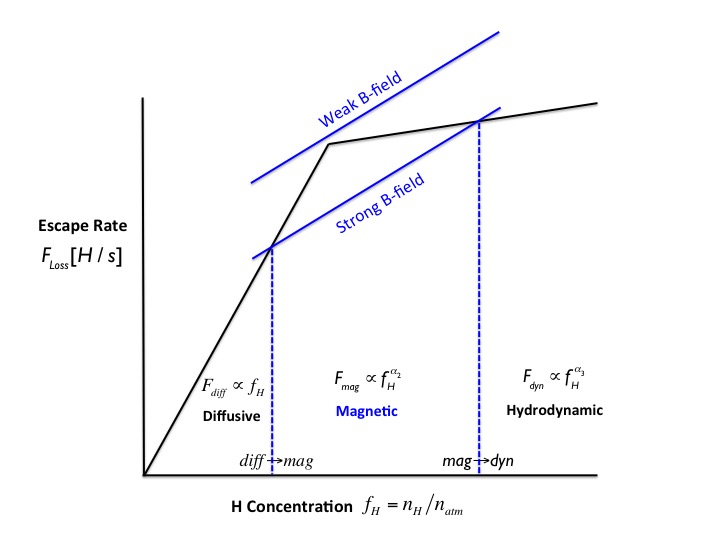}
\caption{Schematic diagram of atmospheric limiting escape regimes.  Hydrogen escape rate versus mixing ratio with the diffusive (black), magnetic (blue), and hydrodynamic (black) regimes.  The transition from diffusive to hydrodynamic escape may be interrupted by magnetically limited escape if the planetary magnetic field is strong.  See \citet{Driscoll2013} for more details.}
\label{fig:escape}
\end{figure}

The hydrodynamic and diffusion limits to escape have been heavily studied \citep[e.g.][]{hunten1973,watson1981,shizgal1996,lammer2008,luger2015b}.  Both of these escape limits depend on hydrogen number density. In the diffusion limit escape is linearly proportional to hydrogen mixing ratio at the tropopause \citep{hunten1973}, while hydrodynamic (or energy) limited escape has a weaker dependence on hydrogen mixing ratio but is expected to occur at much higher H concentrations \citep{watson1981}.  
Hydrodynamic escape is fundamentally driven by incident energy absorbed at the base of the escaping region, but a bottleneck occurs as the escape rate grows and the absorbing species are lost, thereby limiting absorption and the energy available to drive escape.  Regardless of their detailed dependences, diffusion and energy limited escape are expected to intersect at some H mixing ratio (therefore denoting a transition from one mechanism to the other) as depicted in Figure \ref{fig:escape}.

If a strong planetary magnetic field is present that balances the solar wind far from the top of the atmosphere, then the density of species exposed to the flow will be limited.  In other words, the rate at which ionized planetary species are swept away by the stellar wind will decrease with increasing magnetopause distance and magnetic field strength. This effect adds a third limiting escape regime between the diffusion and hydrodynamic regimes.  Figure \ref{fig:escape} illustrates this scenario, where increasing hydrogen concentration leads to a transition from diffusion to magnetic limited, and then from magnetic to hydrodynamic limited escape. \cite{Driscoll2013} propose that the limiting physical mechanism in the magnetic regime could be the removal of planetary ions from the magnetopause via Kelvin-Helmholtz instabilities, the occurrence of which are well documented \citep[e.g.][]{wolff1980,barabash2007,taylor2008}. Mathematically the magnetically limiting escape rate is predicted to be a function of hydrogen ion concentration, magnetopause surface area, and instability time scale, such that stronger magnetic fields expose a lower density of planetary species to the solar wind because their number density decreases faster than the magnetopause surface area \citep{Driscoll2013}.  Weaker (or non-existent) planetary magnetic fields will allow the solar wind to interact with the ``top'' of the atmosphere (where escape occurs), potentially pushing the magnetic limit above the diffusion and hydrodynamic limits, rendering the magnetic limit irrelevant (Figure \ref{fig:escape}).

\subsection{Extent of coupling between mantle, core, and climate}
\label{sec:core_summary}

\begin{figure}
\centering
\includegraphics[width=\linewidth]{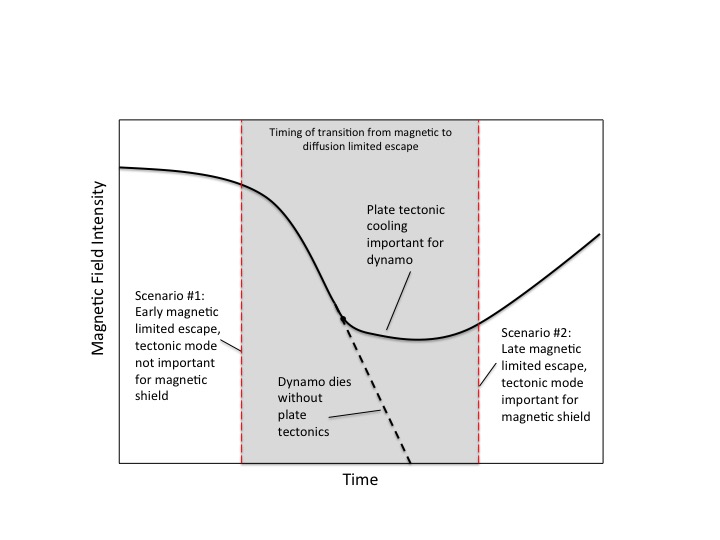}
\caption{\label {fig:magneticregime} Schematic of magnetic field intensity over time.  Initially the dynamo is maintained by cooling during the thermal adjustment period (solid black) and tectonic mode is not important.  This continues until core cooling begins to slow and the dynamo will either die without efficient mantle cooling (dashed black), or stay strong with efficient interior cooling associated with plate tectonics (solid black).  If the timing of this event occurs after the transition from magnetic to diffusion limited escape (Scenario \#1, vertical dashed red) then the shielding of the atmosphere is not dependent on the tectonic mode.  If the dynamo divergence event occurs before the escape transition (Scenario \#2, vertical dashed red) then the shielding of the atmosphere is coupled to the tectonic mode. }
\end{figure}

The magnetic field is therefore most important for atmospheric evolution during the transition from hydrodynamic to diffusion limited escape. A strong magnetic field could reduce the escape rate during this transition, thereby helping to preserve a planet's water budget. Without magnetic shielding significant water loss could occur, potentially preventing the silicate weathering feedback from acting to stabilize climate and, in turn, plate tectonics from operating. Whether plate tectonics is itself necessary for magnetic shielding during the transition to diffusion limited escape depends on the timing of this transition (Figure \ref{fig:magneticregime}). If the transition to the diffusion regime occurs quickly, i.e.\ $f_H$ decreases rapidly, then it would occur during a planet's thermal adjustment period, when dynamo action is possible without plate tectonics (see \S \ref{sec:core_mantle}). On the contrary, if the magnetic limited escape regime is occupied longer than the mantle thermal adjustment period (e.g.\ $f_H$ decreases slowly) then plate tectonics is probably important for preventing significant water loss through magnetic shielding. Unfortunately, the timing of the transition to diffusion limited escape is not well known, as $f_H$ depends on the details of a planet's atmospheric structure. Future work in this area is needed to determine how long rocky planets occupy the magnetic limited escape regime, and the conditions under which plate tectonics is necessary for water retention.

The timing of the transition to diffusion limited escape is also affected by stellar wind strength. In particular, magnetic shielding may be important over a wider range of stratospheric hydrogen mixing ratios, and thus over a longer period of time, for planets exposed to stellar winds much stronger than those at Earth today. Strong stellar winds are likely for planets orbiting active solar mass stars or orbiting close to moderately active small mass stars.  In these cases maintaining a magnetic field may be crucial for preserving liquid surface water over billion year timescales, and such a long lived dynamo likely requires plate tectonics. In fact the magnetic field, climate, and plate tectonics can act as a self-sustaining feedback in this case, where the magnetic field is required to prevent water loss, water is necessary for silicate weathering to keep the climate cool enough for plate tectonics, and plate tectonics is required to drive the dynamo. 

Our knowledge of how planetary magnetic fields influence atmospheric escape is still in its infancy with many unanswered questions. In fact a strong magnetic field may even enhance escape in some instances, by producing a larger interaction cross section with the solar wind, which can concentrate incident energy flux by a factor of $10-100$ \citep{brain2014}. Clearly a more thorough understanding of how the properties of the atmosphere, stellar wind, and magnetic field influence escape rates is needed to constrain the coupling between the core dynamo and climate.

\section{Whole planet coupling and planetary evolution}
\label{sec:summary}

Sections \ref{sec:plate_generation}, \ref{sec:climate_reg_cc}, and \ref{sec:whole_planet} together outline whole planet coupling between climate, plate tectonics, and the magnetic field. In this section we describe how whole planet coupling can potentially explain the Earth-Venus dichotomy, and lead to a number different evolutionary scenarios for rocky exoplanets, many of which would be unfavorable for life. In particular we focus on habitable zone planets, showing how events early in a planet's history, such as the initial atmospheric composition and the timing of the initiation of plate tectonics, play a major role in determining whether long-term habitable conditions can develop.  A planet's volatile inventory is likely important as well. Given our incomplete understanding of plate tectonics, the carbon cycle, the geodynamo, and how they interact, our discussion is mostly qualitative, and only represents some initial hypotheses for the factors governing planetary evolution.

\subsection{The Earth-Venus dichotomy}
\label{sec:Earth_venus}

\begin{figure}[ht!]
\begin{center}
\includegraphics[width=\linewidth]{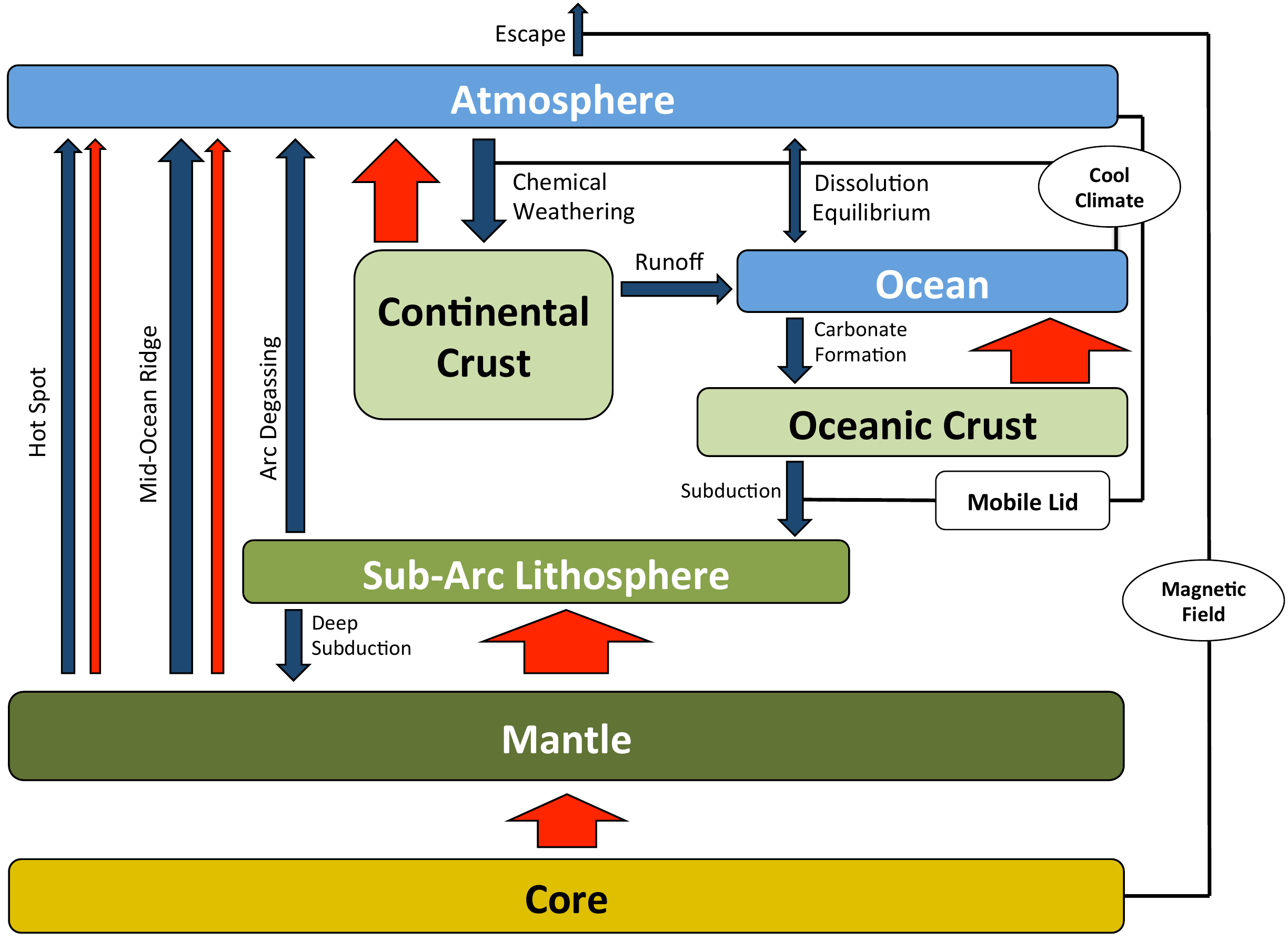}
\caption{Flow chart of climate-tectonic-magnetic coupling for an Earth-like planet. Arrows between reservoirs indicate volatile (blue) and heat (red) fluxes, their width roughly in proportion to magnitude. Black lines indicate conceptual couplings, such as the influence of magnetic field strength on escape rate or chemical weathering on climate. }
\label{fig:flowE}
\end{center}\end{figure}

\begin{figure}[ht!]
\begin{center}
\includegraphics[width=0.6\linewidth]{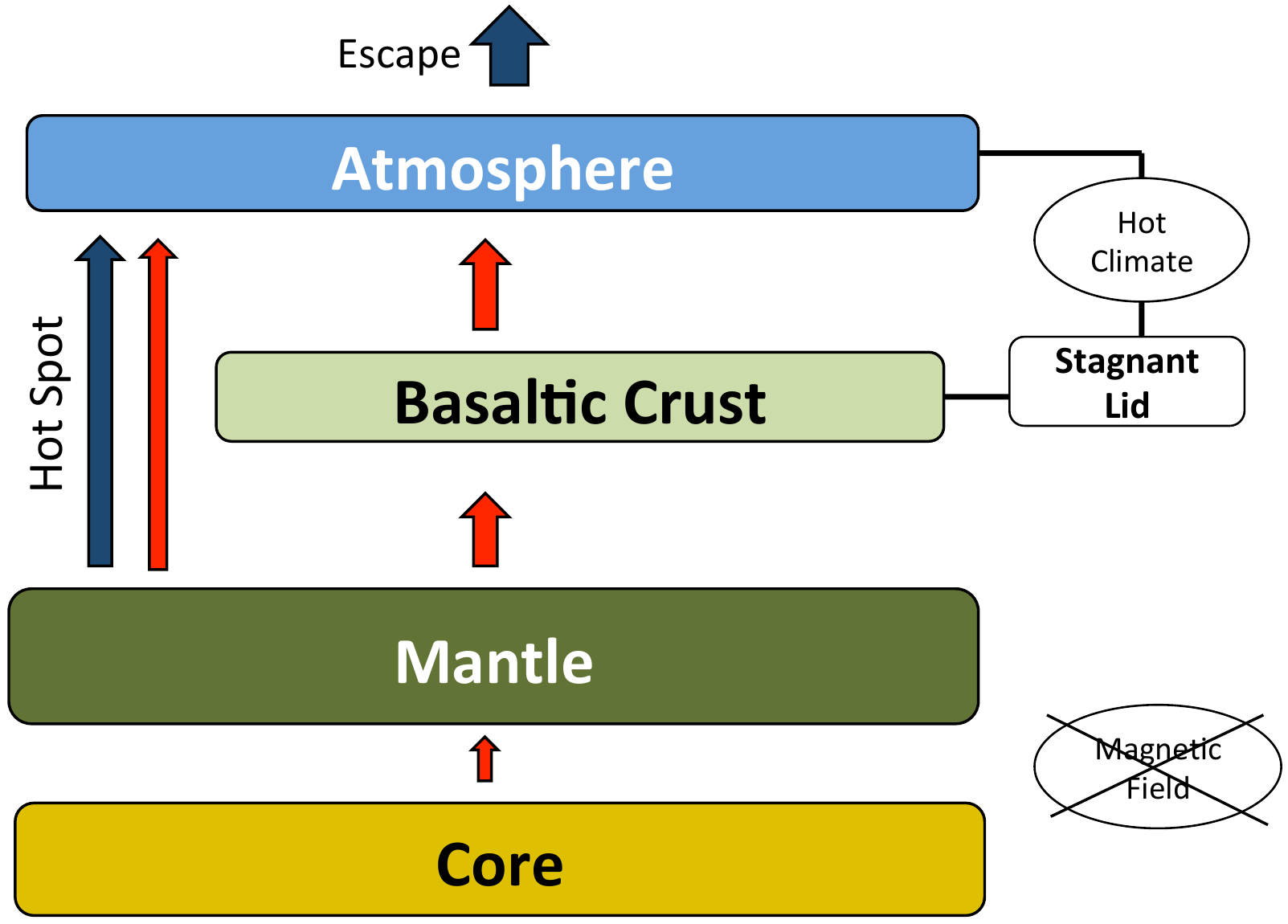}
\caption{Flow chart of climate-tectonic-magnetic coupling on a Venus-like planet. }
\label{fig:flowV}
\end{center}\end{figure}

Figures \ref{fig:flowE} and \ref{fig:flowV} illustrate how we propose that climate, mantle, and the core interact on Earth and Venus.  Venus, being inward of the inner edge of the habitable zone, can not have liquid water at its surface.  As a result, silicate weathering can not draw CO$_2$ out of the atmosphere, so a hot, CO$_2$ rich climate forms via volcanic degassing and/or primordial degassing during accretion and magma ocean solidification. The hot climate prevents plate tectonics, and in turn the long-lived operation of a core dynamo due to a low core heat flux (Figure \ref{fig:flowV}). For the Earth, being in the habitable zone allows liquid water, and in turn silicate weathering. Thanks to a temperate climate plate tectonics can operate, and the resulting high core heat flow powers the geodynamo (Figure \ref{fig:flowE}). \cite{Jellinek2015} invoke a similar series of couplings to explain the Earth-Venus dichotomy, though they argue that the lack of plate tectonics on Venus is caused by high rates of radiogenic heating in the mantle, and that Venus' hot climate stems from the lack of plate tectonics. Their hypothesis contrasts with our interpretation, that Venus' orbital position is the key factor explaining its evolution.  

If surface-interior coupling is responsible for the divergent evolution of Earth and Venus, the runaway greenhouse climate, loss of plate tectonics (if it ever existed), and death of the magnetic field should all be correlated in time. Thus, determining the climate, magnetic, and tectonic history of Venus is a vital test. Unfortunately our current knowledge of Venusian history is poor. The trace amount of water in the atmosphere requires a minimum of about 500 Myr for the H to escape to space \citep{donahue1999}, so the runaway greenhouse must have occurred by at least $\approx 0.5-1$ Ga, but could have taken place much earlier. In fact, Venus may have entered a runaway greenhouse and lost its water during formation \citep{Hamano2013}.  

The Venusian tectonic and magnetic histories are similarly uncertain. There is evidence for a massive resurfacing event at 0.5-1 Ga. Surface features related to this event are most consistent with volcanism and lava flows rather than subduction of old crust and creation of new crust at ridges \citep{smrekar2013,ivanov2013}.  The planet may still be volcanically ``active'' today \citep{smrekar2010}, but eruptions are likely sporadic.  
The style of tectonics before the resurfacing event is unknown, with stagnant lid convection, episodic overturns, or even Earth-like plate tectonics all possibilities. Unraveling the Venusian magnetic history is challenging because the preservation and in-situ measurement of any remanent magnetization on Venus' hot surface is unlikely. However, another possible line of evidence that may imply the presence of a paleo-Venusian magnetic field would be the loss of H$^+$, He$^+$, and O$^+$ along polar magnetic field lines to the solar wind, a process known on Earth as the polar wind \citep{moore2007}.  This ion escape mechanism relies on the presence of an internally generated magnetic field, and could potentially leave a chemical fingerprint in the Venusian atmosphere \citep{brain2014}. However our present day knowledge provides no evidence for a strong, internally generated magnetic field on Venus. Future exploration of Venus is needed to place tighter constraints on its evolution.   

\subsection{Evolution of rocky exoplanets}
\label{sec:sum_exoplanets}

Venus demonstrates one likely evolutionary scenario for planets lying inward of the habitable zone's inner edge. However, cooler climates, that still allow plate tectonics and a magnetic field, are also possible for such planets if they have a significantly smaller CO$_2$ inventory than Venus.  In this case a temperate climate can still exist, even without liquid water and silicate weathering to regulate atmospheric CO$_2$ levels, simply because there isn't enough CO$_2$ to cause extreme greenhouse warming. In some cases a planet that experiences a runaway greenhouse could even retain some water at the poles, and potentially remain habitable \citep{Kodama2015}. Planets lying beyond the habitable zone's outer edge will likely be cold, and thus can plausibly sustain plate tectonics and a magnetic field as well. However, whether complex life can develop on any of these planets is unclear.    

Likewise, an Earth-like evolution is one likely scenario for planets lying within the habitable zone, but a number of factors could lead to a different evolution that is unsuitable for life. Hot, CO$_2$ rich climates are expected after planet formation due to degassing during planetesimal accretion and magma ocean solidification \citep{Abe1985,Zahnle2007,Lindy2008}. If a planet's initial climate is so hot that liquid water is not stable (i.e. temperature and pressure conditions are beyond the critical point for water), then developing a temperate climate is probably not possible. Initial surface temperature and pressure conditions in excess of the critical point imply that silicate weathering would not occur, with or without plate tectonics (though some limited reaction between atmospheric CO$_2$ and the crust is possible).  As a result the climate would remain extremely hot, preventing both plate tectonics and a core dynamo. However, a climate hot enough to exceed the water critical point is an extreme case, requiring $\approx 500$ bar of CO$_2$ for a planet with an atmosphere composed of CO$_2$ and H$_2$O and receiving the same insolation as the Hadean Earth \citep{Lebrun2013}. Earth's total planetary CO$_2$ budget is estimated at $\approx 100-200$ bar, so even with complete degassing during accretion or magma ocean solidification liquid water was still stable \citep[e.g.][]{Sleep2001b,Zahnle2007}. Thus a planet would need a significantly larger total CO$_2$ budget than Earth for initial atmospheric makeup to exceed the liquid water critical point.   

\begin{figure}
\centering
\includegraphics[scale = 0.6]{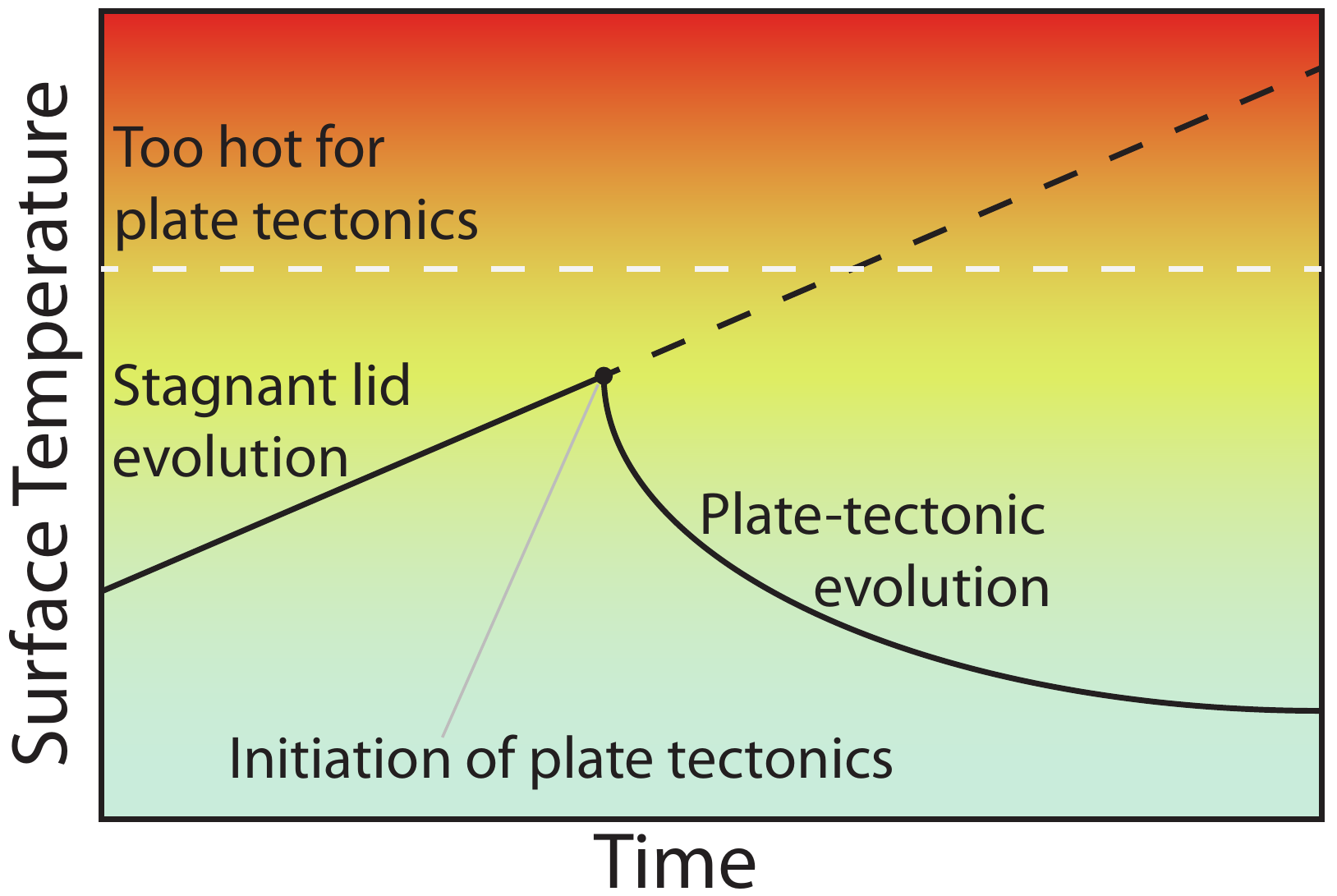}
\caption{\label {fig:div_point} Schematic diagram of the divergence point in planetary evolution involving the initiation of plate tectonics. Before plate tectonics a planet's weathering rate may be supply limited, such that surface temperature climbs over time as a result of increasing luminosity. Once plate tectonics initiates, silicate weathering is enhanced by higher erosion rates and continent formation, and climate cools. }  
\end{figure}

Another possibility is a climate where liquid water is stable, but is still too hot for plate tectonics. In this case low land fractions or erosion rates can potentially prevent plate tectonics, and a cool climate, from ever developing. High rates of volcanism would be needed to avoid this fate, by creating a sufficient supply of fresh rock at the surface for silicate weathering to cool the climate. Finally, even when climate conditions are amenable to plate tectonics, an uninhabitable state can be reached if plate tectonics does not initiate before increasing insolation warms the climate to the point where it is no longer possible (Figure \ref{fig:div_point}). Before plate tectonics initiates on a planet, weathering may be supply limited and the atmosphere CO$_2$ rich. Thus surface temperature will increase with increasing luminosity, and can potentially become hot enough to preclude plate tectonics from ever starting. If plate tectonics does not initiate before this divergence point is reached, a planet could become permanently stuck with a hot, uninhabitable climate, stagnant lid convection in the mantle, and no protective magnetic field. 

Conversely on a planet where plate tectonics does initiate before reaching this divergence point, continental growth and orogeny will increase both land area and erosion rates, enhancing the ability of silicate weathering to establish and maintain a temperate, habitable climate. If a large enough land area forms, weathering may even be capable of maintaining a temperate climate without plate tectonics to elevate erosion rates. Likewise plate tectonics means that long-lived dynamo action, and therefore volatile shielding from the solar wind, is possible. The divergence point involving the initiation of plate tectonics could be particularly important if high mantle temperatures are a significant impediment to plate tectonics through low convective stresses or the creation of thick buoyant crust (see \S \ref{sec:pt_other_factors}). Mantle temperature may need to cool before plate tectonics can begin, even if surface temperatures are not hot enough to preclude plate motions. However, the increase in luminosity during a star's main sequence evolution is relatively gradual (the sun's luminosity was only $\approx 30$ \% lower 4.5 Gyrs ago \citep{Gough1981}), so there is a large time window, on the order of 1 Gyr, for sufficient mantle cooling to occur before climate becomes too hot for plate tectonics. 
 
 Another divergence point involves magnetic shielding and planetary water loss. The magnetic field can limit H escape, and thus help preserve surface water, when atmospheric escape transitions from being hydrodynamically limited to diffusion limited (see \S \ref{sec:mag_limited_escape} and \S \ref{sec:core_summary}.) If no magnetic field is available to shield the solar wind then massive amounts of H may be lost, leaving the planet desiccated and unable to regulate atmospheric CO$_2$ levels. As discussed in \S \ref{sec:core_summary}, if the transition to diffusion limited escape is early in a planet's history, then magnetic shielding is possible regardless of tectonic regime, and many planets will likely be able to keep their volatiles and possibly develop habitable climates. However, if the transition occurs after the mantle's thermal adjustment period, then plate tectonics is likely necessary for magnetic shielding, and fewer planets will be able to retain surface water.

\begin{figure}
\includegraphics[scale = 0.65]{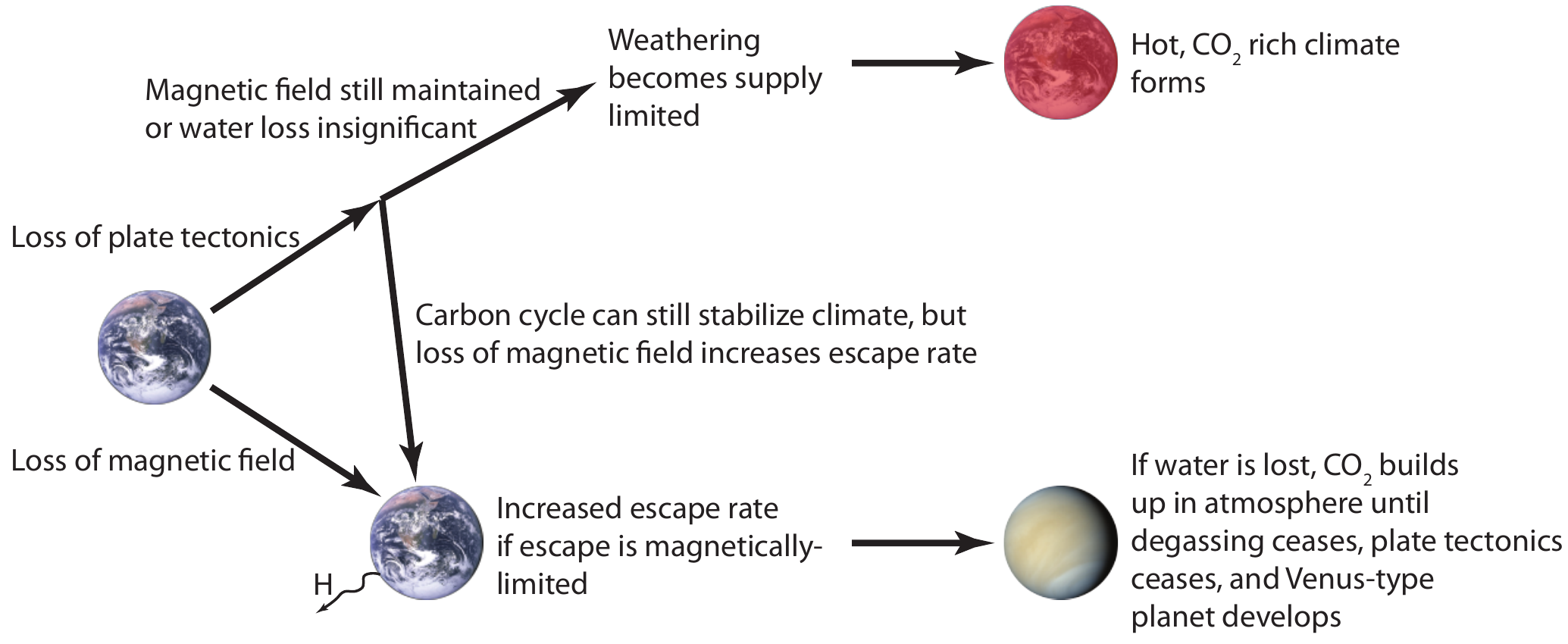}
\caption{\label {fig:divergent_evol} Schematic diagram of possible scenarios where failure of the feedbacks between climate, plate tectonics, and the magnetic field leads to divergent evolution of terrestrial planets within the habitable zone. }  
\end{figure}

Additional divergence points are possible later in planetary evolution if climate, tectonics, and the magnetic field are tightly coupled. For example, if the carbon cycle and plate tectonics act as a self-sustaining feedback, where high erosion rates, supplied by plate tectonics, are required to maintain kinetically limited weathering and thus keep surface temperatures cool enough for plate tectonics (see \S \ref{sec:pt_cc}), then the loss of plate tectonics would lead directly to a hot climate state that likely precludes plate tectonics from re-initiating. Without plate tectonics to enhance erosion, the inhospitably hot climate that results from supply limited weathering would likely be permanent. Such tight coupling between plate tectonics and the carbon cycle is most likely on planets with small exposed land areas or high total CO$_2$ budgets. Another divergence point is possible for planets orbiting very active solar mass stars or very close to active small mass stars where stronger stellar winds can more efficiently strip atmospheric volatiles. For these planets cessation of the core dynamo could cause significant water loss, in turn halting silicate weathering and, due to the ensuing hot climate, likely shutting down plate tectonics as well (see Figure \ref{fig:divergent_evol} for a schematic illustration of the divergence points discussed in this paragraph).  

Size is also a major factor in terrestrial planet evolution.  Large planets are expected to have wider habitable zones due to the influence of higher gravity on atmospheric scale height and the greenhouse effect \citep{Kopp2014}, but also shallower ocean basins and thus less exposed land, unless feedbacks can regulate ocean volume such that continents are always exposed \citep[][ see also \S \ref{sec:future}]{Cowan2014}. The influence of size on plate tectonics and magnetic field generation is also unclear. Previous studies have found that plate tectonics is more likely on larger planets \citep{Valencia2007b,Valencia2009,vanheck2011,Foley2012}, less likely on larger planets \citep{ONeill2007,Kite2009,Stein2013,noack2014,stamenkovic2014,miyagoshi2014,tachinami2014}, or that size is relatively unimportant \citep{Korenaga2010a}. Different studies reach very different conclusions because of the large uncertainties in the rheological mechanism necessary for generating plate tectonics, how key features such as internal heating rate scale with size, and how mantle properties are affected by pressure and temperature. The influence of size on magnetic field strength is likewise debatable.  Several studies have found a weak dependence of field strength and lifetime on planet and core size, and possibly a peak in strength for Earth-sized planets \citep{gaidos2010,tachinami2011,driscoll2011a,vansummeren2013}. Generally, larger dynamo regions are expected to produce stronger magnetic fields \citep[e.g.][]{christensen2009}, but variations in more subtle properties, like mantle and core composition, likely play a fundamental role. The coupling between plate tectonics, climate, and the magnetic field is expected to apply to planets of different size, but future work is needed to place tighter constraints on the influence of size on this coupling and on planetary evolution.  

The issue of size highlights the many uncertainties remaining in our knowledge of planetary dynamics and evolution.  Each aspect of planetary dynamics that is important for magnetic, tectonic, and climate evolution needs to be better understood before more rigorous predictions or interpretations can be made.  In particular the interactions between different components of the planetary system, specifically interactions between surface tectonics, mantle convection, and the long-term carbon cycle, between mantle convection and the core dynamo, and between the magnetic field, atmospheric escape, and climate, deserve significant attention.  With a large number of rocky exoplanets already discovered, many of which are in their respective habitable zones \citep{Batalha2014}, and more certain to follow, improving our knowledge of planetary evolution is an important goal. 

\section{Future Directions} 
\label{sec:future}

In addition to furthering our understanding of plate tectonics, magnetic field generation, climate evolution, and the interactions between these processes, there are a number of new questions and research topics that must be addressed to advance our knowledge of planetary evolution. We summarize a few of these important questions below.  

1) {\bf What are the material properties of Earth-like and non-Earth-like terrestrial planets at high temperature, high pressure conditions?} Without tighter constraints on basic properties like density, viscosity, and thermal conductivity models of exoplanet evolution will continue to be highly uncertain. Conducting both laboratory and numerical experiments at the conditions relevant for Earth and super-Earths is challenging, but is also vital for determining how terrestrial planets behave. Moreover, the composition of exoplanets could differ significantly from Earth, so constraints on the properties of non-Earth-like materials are necessary for modeling the evolution of these planets as well.  

2) {\bf What controls the volume of water at a planet's surface (i.e. the size of the oceans and the amount of subaerial land)?}  Earth has maintained a relatively constant freeboard, or water level relative to the continents, throughout much of its history \citep{Wise1974}, despite the fact that there could be multiple oceans worth of water stored in the mantle \citep[e.g.][]{Ohtani2005}. Are there feedbacks in Earth's deep water cycle acting to keep a large continental surface area exposed, as proposed by some \citep{Kasting1992,Cowan2014}? Given the importance of exposed land to the long-term carbon cycle, this question is of fundamental importance to planetary evolution. 

3) {\bf What are typical volatile abundances, in particular water and carbon dioxide, for rocky planets, and how much of this volatile inventory resides in the atmosphere just after accretion and magma ocean solidification?} Both volatile inventory and initial atmospheric composition are important for planetary evolution. Thus, a better understanding of how and when volatiles are delivered during accretion, how much degassing occurs during accretion and solidification of a possible magma ocean, and how much atmospheric loss occurs during this early phase of planetary history, are all necessary for constraining planetary evolution. 

4) {\bf Can strong gravitational tides render a planet uninhabitable?}  Planets experiencing strong gravitational tides (caused by nearby stars, planets, or satellites) can generate significant internal heating via tidal dissipation, which can cause extreme surface volcanism and hinder dynamo action \citep{driscoll2015b}. Efficient cooling of the interior could allow the orbits of such planets to circularize faster, minimizing the length of time spent in a tidally heated regime, and could move the planet in (or out) of the habitable zone. Future work should explore the details of tidal dissipation in the mantle and core, and how tides can influence long-term evolution.

5) {\bf How and when does plate tectonics initiate?} Factors other than climate, like mantle temperature, can have an important control over whether plate tectonics can operate.  Thus even with favorable climate conditions, other process such as mantle cooling may be necessary for Earth-like plate tectonics. Understanding the type of tectonics that might take place on a planet before plate tectonics, how much land and weatherable rock can be created through non-plate-tectonic volcanism, and the factors that then allow for Earth-like plate tectonics to develop, will all be crucial for determining how likely planets are to follow an evolutionary trajectory similar to Earth's.


%
%
%
%
%
%

%
%
%
%

\section{Acknowledgements}
BF acknowledges funding from the NASA Astrobiology Institute under cooperative agreement NNA09DA81A. We thank Norm Sleep for a thorough and constructive review that helped to significantly improve the manuscript. No new data was used in producing this paper; all information shown is either obtained by solving the equations presented in the paper or is included in papers cited and listed in the references. 

%
%
%
%
%
%
%
%
%
%

\end{document}